\documentclass[lettersize,journal]{IEEEtran}
\usepackage{amsmath,amsfonts}
\usepackage{algorithmic}
\usepackage{algorithm}
\usepackage{array}
\usepackage[caption=false,font=normalsize,labelfont=sf,textfont=sf]{subfig}
\usepackage{textcomp}
\usepackage{stfloats}
\usepackage{url}
\usepackage{verbatim}
\usepackage{graphicx}
\usepackage{cite}

\usepackage{color}
\usepackage{booktabs}
\usepackage{multirow}

\begin{document}

\title{A Unified Generative Framework based on Prompt Learning for Various Information Extraction Tasks}

\author{Zhigang Kan, Linhui Feng, Zhangyue Yin, Linbo Qiao$^{*}$, Xipeng Qiu, Dongsheng Li$^{*}$
        % <-this % stops a space

\thanks{$^{*}$ corresponding author.}

\thanks{
Zhigang Kan, Linhui Feng, Linbo Qiao, and Dongsheng Li are with the School of Computer, National University of Defense Technology, Changsha, China. Email: \{kanzhigang13, fenglinhui19, qiao.linbo, dsli\}@nudt.edu.cn.}
\thanks{Zhangyue Yin and Xipeng Qiu are with the School of Computer Science, Fudan University. Email: yinzy21@m.fudan.edu.cn, xpqiu@fudan.edu.cn.}
% \thanks{This work was supported by the National Natural Science Foundation of China under Grant No. 62025208 and the Open Fund of Science and Technology on Parallel and Distributed Processing Laboratory (PDL).
% Part of this work was done when the first author was visiting Fudan University.}% <-this % stops a space

% \thanks{\begin{itemize}
% \item Zhigang Kan, Linhui Feng, Linbo Qiao, and Dongsheng Li are with the School of Computer, National University of Defense Technology, Changsha, China.\\
% \item Zhangyue Yin and Xipeng Qiu are with the School of Computer Science, Fudan University. \\
% Email:yinzy21@m.fudan.edu.cn, xpqiu@fudan.edu.cn.
% \end{itemize}}% <-this % stops a space

% \thanks{Manuscript received September xx, 2022; revised September xx, 2022.}
}

% \markboth{IEEE Transactions on Neural Networks and Learning Systems,~Vol.~00, No.~00, September~2022}%


\maketitle

\begin{abstract}
Prompt learning is an effective paradigm that bridges gaps between the pre-training tasks and the corresponding downstream applications. 
Approaches based on this paradigm have achieved great transcendent results in various applications.
However, it still needs to be answered how to design a unified framework based on the prompt learning paradigm for various information extraction tasks.
In this paper, we propose a novel composable prompt-based generative framework, which could be applied to a wide range of tasks in the field of Information Extraction. 
Specifically, we reformulate information extraction tasks into the form of filling slots in pre-designed type-specific prompts, which consist of one or multiple sub-prompts.
A strategy of constructing composable prompts is proposed to enhance the generalization ability to extract events in data-scarce scenarios.
Furthermore, to fit this framework, we transform Relation Extraction into the task of determining semantic consistency in prompts.
The experimental results demonstrate that our approach surpasses compared baselines on real-world datasets in data-abundant and data-scarce scenarios.
Further analysis of the proposed framework is presented, as well as numerical experiments conducted to investigate impact factors of performance on various tasks.
\end{abstract}


\section{Introduction}

\IEEEPARstart{W}ith the development of deep learning and the application of distributed technology in training large-scale models, Pre-trained Models have gradually emerged in many fields such as Computer Vision (CV) and Natural Language Processing (NLP). 
Along with Pre-trained Models came a new research paradigm—``fine-tuning''. It is the pattern that pre-trains models on large-scale unlabeled data to learn common knowledge first, then fine-tunes them for downstream tasks.
Benefitting from the comprehensive knowledge learned during the pre-training phase, Pre-trained Models could achieve considerable performance on downstream tasks.

\begin{figure}[t]
\vspace{-1em}
\centering
\includegraphics[width=1\columnwidth]{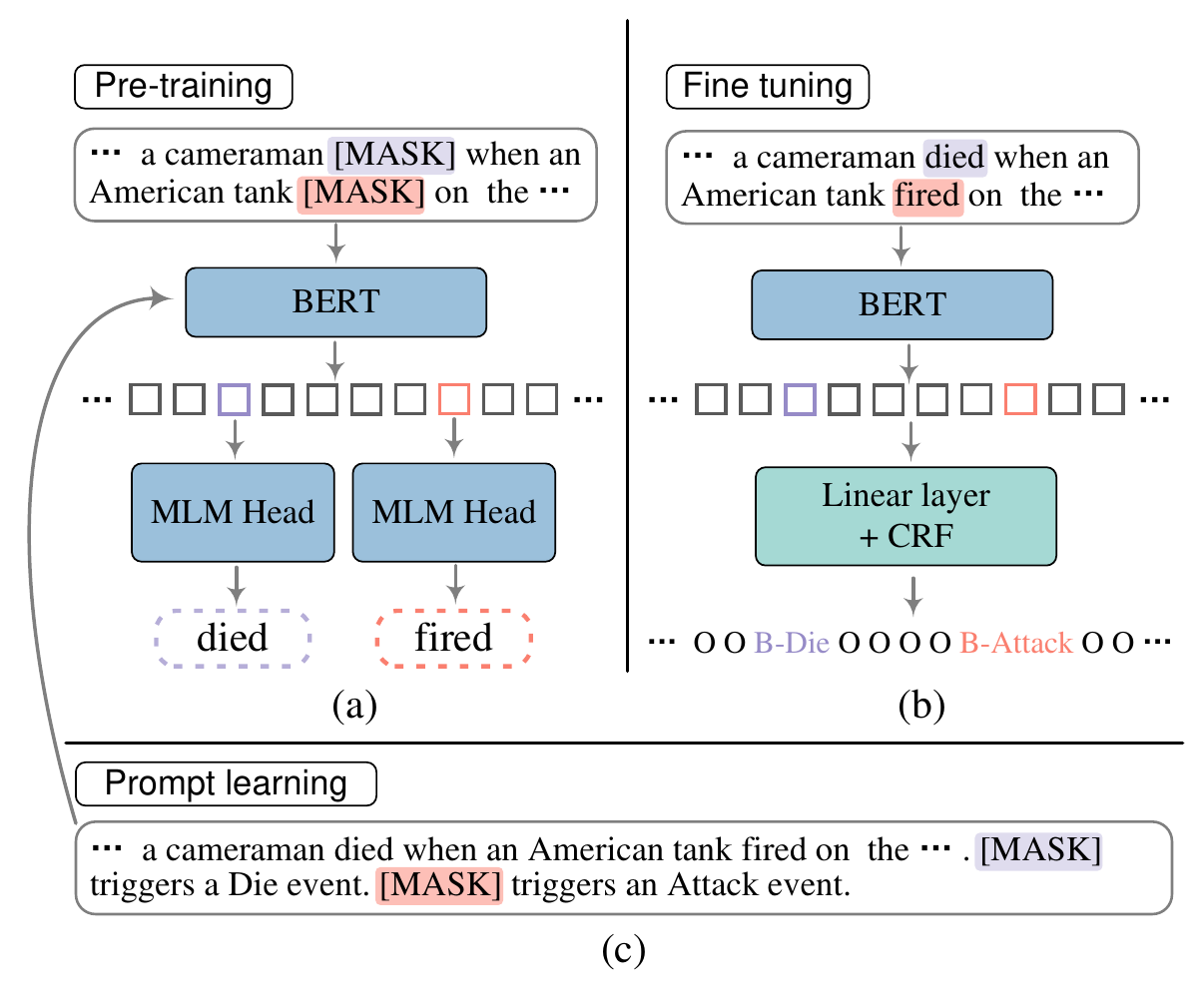} 
%\vspace{-1em}% Reduce the figure size so that it is slightly narrower than the column. Don't use precise values for figure width.This setup will avoid overfull boxes.
% \caption{An example of pre-training and fine-tuning processes. The upper part shows the pre-training process of BERT, and the lower part is the overview of an approach that detects events by applying the CRF algorithm and fine-tuning BERT.}
\vspace{-1em}
\caption{Processes of pre-training BERT and applying it based on the ``fine-tuning'' and ``prompt learning'' paradigm. (a) is the training diagram of the ``masked language model'' task, which is the most representative sub-task in the BERT pre-training process. (b) is the overview of a ``fine-tuning''-based approach that detects triggers by applying the CRF algorithm and fine-tuning BERT. (c) shows how to construct the input for a ``prompt learning''-based method to detect triggers. The network structure of this method is the same as the ``masked language model'' sub-task in pre-training.}

\label{pretrain_finetuning_prompt_compare}
\end{figure}

Pre-trained Models for natural language are also known as Pre-trained Language Models (PLMs), which brought new vitality to the NLP field due to the remarkable ability to express text.
The most famous case is BERT \cite{BERT_2018}, which achieved the best results on 11 tasks at the beginning of its publication and was later introduced into various NLP fields.
Since the emergence of BERT, how to exploit the potential of PLMs for downstream tasks has been an important research point.

In the field of Information Extraction (IE), traditional methods \cite{Multi_grained_NER_2019,PLMEE_Yang_2019,fine_tuning_RE_1} are based on the fine-tuning paradigm.
They construct models by attaching well-designed networks for the downstream tasks to the tail of PLMs. 
% And then, these methods iteratively update the parameters of PLMs and additional networks with the help of supervised information during the training process.
% A series of valuable works \cite{Multi_grained_NER_2019,PLMEE_Yang_2019,fine_tuning_RE_1} have demonstrated the effectiveness of this paradigm.
However, as shown in Fig.\ref{pretrain_finetuning_prompt_compare}, at least two factors lead to a non-negligible gap between pre-training and fine-tuning when applying this paradigm.
First, applying extra networks leads to structural differences between pre-training and fine-tuning.
Secondly, the differences between pre-training and downstream tasks result in a significant discrepancy.
% In early years, when dealing with tasks of information extraction, researchers usually attach a well-designed network for the downstream task to the tail of a Pre-trained Language Model.
% It 
    
\begin{figure*}[thb]
\vspace{-1em}
\centering
\includegraphics[width=1.98\columnwidth]{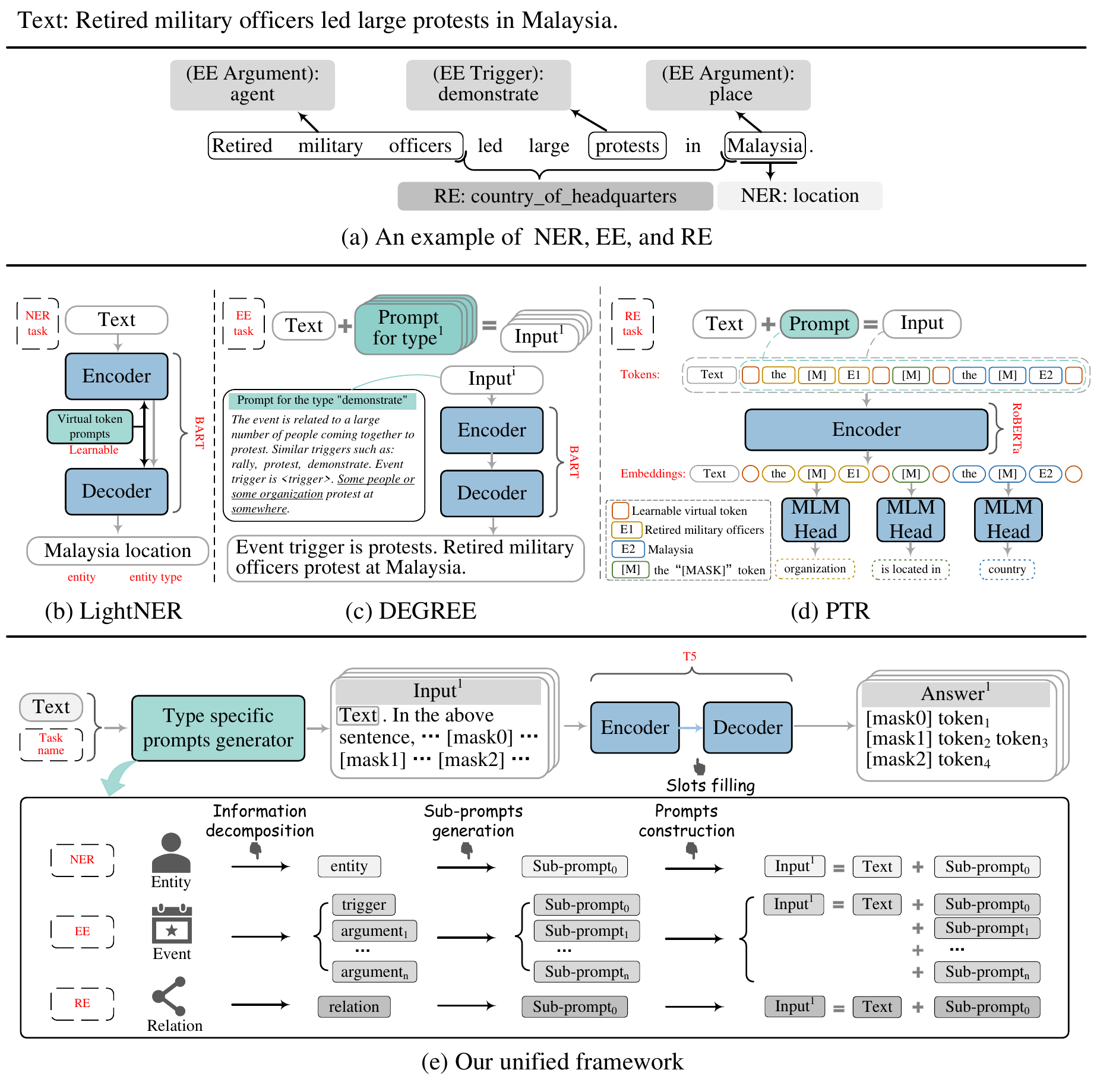} 
%\vspace{-1em}% Reduce the figure size so that it is slightly narrower than the column. Don't use precise values for figure width.This setup will avoid overfull boxes.
% \vspace{-1em}
% \vspace{-1em}
\caption{Overviews of existing classic prompt-based frameworks for IE tasks and our unified composable prompt-based generative framework. (a) is an example of extracting entities, events, and relations from a sentence. (b) is the overview of LightNER, a prompt-based method for the NER task. (c) shows the framework of DEGREE, which completes the EE task by recovering damaged text in every type-specific prompt. (d) displays the details of extracting relation via the RoBERTa-based approach PTR. (e) is an example of using the framework proposed in this paper to accomplish the NER, EE, and RE tasks uniformly. }
\label{difference_from_current_methods}
% \vspace{-1em}
\end{figure*}

% One of the these tremendous achievements is BERT\cite{BERT_2018}, which introduced two classic self-supervised tasks: Masked Language Model (MLM) and Next Sentence Prediction (NSP) for PLMs and obtained the best results on eleven NLP tasks.
% The success of BERT brought new vitality to the field of NLP. 
% On the one hand, researchers have continuously proposed various PLMs by optimizing BERT \cite{RoBERTa_2019} or designing new architectures \cite{BART_2020} and applied them to refresh the performance rankings of major NLP tasks. 
% On the other hand, Academia and industry are also working on exploiting the potential of PLMs more effectively on downstream tasks \cite{what_happen_2020}.
% In applying PLMs, fine-tuning pre-trained models according to the characteristics of target tasks is promising.
% In early years, researchers in the filed of Information Extraction (IE) 

% Methods based on the ``fine-tuning'' paradigm usually attach a well-designed network for the downstream task to the tail of a Pre-trained Language Model and iteratively update the parameters of the PLM and that network with the help of supervised information during the training process.
% A series of valuable works have demonstrated the effectiveness of this paradigm \cite{PLMEE_Yang_2019}.

Recently, a new paradigm called ``prompt learning'' was proposed to bridge gaps between fine-tuning and pre-training \cite{knowledge_2019,Gao_2021}.
Same as the example in Fig.\ref{pretrain_finetuning_prompt_compare}(c), in this paradigm, various downstream tasks are converted to tasks (such as cloze) that are familiar to PLMs.
Based on the prompt learning paradigm,  Chen et al. proposed a generative framework named ``LightNER'' \cite{lightNER} for the Named Entity Recognition (NER) task.
The overview of the method is shown in Fig.\ref{difference_from_current_methods}(b).
It introduced a prompt-guided attention layer that utilizes learnable prompts to guide the focus of attention.
This framework achieves encouraging results on low-resource NER tasks but cannot apply to other IE tasks such as Event Extraction (EE).

CondiGen \cite{CondiGen_Li_2021} is an event argument extraction model based on the prompt learning paradigm, which extracts arguments by utilizing the BART \cite{BART_2020} to recover the damaged template of each event type. It did not answer how to extract triggers by the prompt learning paradigm and suffers from error propagation due to the pipeline structure.
In 2021, Hsu et al. proposed ``DEGREE'', a prompt-based generative model, which provides a template design approach for EE \cite{degree}.
Specifically, it extracts events by restoring damaged texts in type-specific prompts.
Fig.\ref{difference_from_current_methods}(c) is the schematic diagram of this approach with joint structure.
DEGREE achieves encouraging results in both low- and high-resource scenarios.

% However, the prompt templates proposed in DEGREE do not cover all argument roles in the corpus ACE 2005, which is used as the dataset in their experiments.
% According to our statistics, 67\% of the event types in their templates lost some argument roles, such as the type ``End-Org'' misses the role ``Place''.
% Even 18.5\% of argument roles (such as "Crime") do not appear in any events.
% Furthermore, DEGREE constructs the prompt for arguments in an event by gathering all pronouns of the argument roles into a short sentence that describes the occurrence of that event.
However, DEGREE constructs the type-specific prompt for an event type by gathering all pronouns of the argument roles into a short sentence describing the event's occurrence.
It results in a close correlation between the argument sub-prompt and the event type. 
In other words, this leads to considerable differences between argument sub-prompts for different event types.
For instance, although both the ``Meet'' and ``Demonstrate'' events have the same ``Entity'' and ``Place'' roles, their argument sub-prompts are semantically far apart in DEGREE.
In this situation, DEGREE cannot learn about arguments of ``Meet'' from similar event types with the same roles such as ``Demonstrate''.
Also, DEGREE is not applicable to other IE tasks, such as Relation Extraction (RE).
% DEGREE's templates can't cover all event arguments in the corpus ACE 2005, which is used as the dataset in their experiments.

Prompt Tuning with Rules (PTR) \cite{PTR_2021} is a general prompt-based framework for text classification.
The workflow of PTR is displayed in Fig.\ref{difference_from_current_methods} (d).
It generates prompts by combining sub-prompts and learnable virtual tokens and judges the type of relationship by mapping the ``[MASK]'' tokens to candidate answers via encoder-only PLMs (such as RoBERTa \cite{RoBERTa_2019}). 
% It proposed a general framework for text classification with pr`e-designed sub-prompts and logic rules.
However, PTR can only classify specific text content with pre-designed sub-prompts and logic rules but cannot extract information with a variable length, such as entities and events from texts.
% Moreover, the framework of the PTR prevents it from handling more complex situations where there are more than two relationships between entities.
% In addition, this approach independently predicts each masked word in a multi-slot prompt without exploiting potential associations of masked information fragments.
% Another quality research is Template-based BART \cite{prompt_ner_BART} which extracts named entities by enumerating all possible spans and taking the span with the largest possibility as the final entity.
% But, this method is not designed for more complex tasks such as Event Extraction, which aims to extract structured event information consisting of a trigger and a certain number of arguments.

To mitigate these issues, we introduce a novel unified Composable Prompt-based Generative Framework (CPGF) for all Information Extraction tasks.
As shown in Fig.\ref{difference_from_current_methods} (e), CPGF converts various Information Extraction tasks into the cloze task and applies a generative language model T5 \cite{T5_Raffel_2020} to predict answers.
In order to effectively extract complex information with multiple elements (such as events), we independently design a slot-filling sub-prompt for each component of target information.
Explicitly speaking, CPGF firstly chooses a short describing sentence or the definition in the annotation guidelines as the stem of the cloze task for each element (such as an argument role in the EE task) in target information.
And then, it replaces the keywords in the stems with ``[mask]'' tokens to construct independent sub-prompts for information elements.
Finally, CPGF obtains each type-specific prompt by connecting sub-prompts related to the elements that appear in the target information type.

To enhance the generalization ability of our framework on complex tasks, we further design modular sub-prompts to generate composable prompts.
In detail, we build a modular sub-prompt library by constructing a common sub-prompt for each cluster of elements that share the same meaning but distribute across different types of information.
% , such as the ``Entity'' and ``Place'' in Fig.\ref{prompt_compare}
Each composable prompt is obtained by combining one or multiple modular sub-prompts selected from the library.
In this case, when extracting information with types never seen during training, our framework could effectively extract elements that have appeared in the training set. 
% Therefore, The design of modular sub-prompts enhances the transferability of our framework.

Consider that the Relation Extraction task aims to assign relation types to entity pairs rather than extracting spans from given text. 
Therefore, we carefully design a novel prompt template for this task.
CPGF achieves the purpose of assigning relation types by predicting the semantic contradiction between the original text and prompts.

% In the face of uncertain samples, in which the model makes mistakes, it is difficult for the model to learn the knowledge with a single modular sub-prompt when training for some complex task, e.g. Event Extraction.
% For this reason, we propose a Word Hint strategy to adjust the sub-prompt dynamically.
% In detail, whenever the model mispredicts an answer for a sub-prompt, we give the first word of the correct span and let the model predict the rest during the next epoch.

To verify the effectiveness of the framework proposed in this paper, we conduct numerical experiments on Named Entity Recognition, Event Extraction, and Relation Extraction, which are classic sub-tasks in Information Extraction. In summary, the contributions of this work are four-fold:
\begin{itemize}
% \item We propose a novel unified generative framework based on the prompt paradigm for Information Extraction and a method of constructing prompts for complex information with independent sub-prompts.  
\item A novel unified generative framework based on the prompt learning paradigm for various Information Extraction tasks is proposed.
\item We introduce a method of constructing prompts with independent sub-prompts for complex tasks such as Event Extraction. Furthermore, to improve our framework's generalization ability to extract events in data-scarce scenarios, we design a kind of composable prompts consisting of multiple modular sub-prompts for Event Extraction.
% \item We propose a novel template for Relation Extraction that transforms the task into judging semantic contradictions.
\item We propose a prompt-based method for Relation Extraction by judging semantic contradictions and design a corresponding template for it.
\item A series of experiment results on Event Extraction, Named Entity Recognition, and Relation Extraction demonstrate the effectiveness of our framework in both data-abundant and data-scarce scenarios.
\end{itemize}

\section{Related Works}

\subsection{Prompt Learning Paradigm}

% Methods based on the ``prompt-learning'' paradigm reformulate downstream tasks into a PLM's pre-training task instead of adapting language models to downstream tasks \cite{prompt_liupengfei}.
% In practice, the researchers would design templates with empty slots and train the model to fill in slots with appropriate words or phrases.
Researches \cite{do_prompt_, prompt_shunxu} show that the prompt template's quality dramatically affects the performance of prompt-based models.
The most intuitive way is to manually design a reasonable template for every possible situation based on the human experience. 
Schick and Schütze applied manual templates to the text classification and conditional text generation tasks in the few-shot scenario \cite{Schick_1}.
%However, the efficiency of manually designed templates depends on the designer's experience and a lot of experimentation, which makes it not an easy task to complete.
%Therefore, many approaches are dedicated to obtaining high-quality templates automatically.
Some works are devoted to automatically searching for suitable templates made up of natural language by approaches named Prompt Mining \cite{jiang_2020}, Prompt Paraphrasing \cite{yuan_2021}, Prompt Generation \cite{Gao_2021}, and Prompt Scoring \cite{Davison_2019}.
Other studies believe that applying prompts is intended to guide the model to predict the desired answer.
Therefore, the prompt can be in other forms that machines can understand and is not limited to human language.
For example, Shin et al. fine-tuned embeddings of virtual tokens \cite{Shin_2020}, which are initialized by a discrete search method, as a soft prompt for PLM.
Hard-Soft Prompt Hybrid Tuning is a type of work between the above two directions.
It adds learnable vectors at the beginning and end of the artificially designed prompts to enhance their presentation ability \cite{PTR_2021}.
% Combining the methods above with Information Extraction would be a promising research direction.

\subsection{Information Extraction}

Named Entity Recognition, Event Extraction, and Relation Extraction are three essential sub-tasks of Information Extraction.
% New methods for them are explored every year.
With the recent advances in transformer-based PLMs, many elaborate methods, which apply PLM as the backbone, have been proposed for these three sub-tasks \cite{knowprompt_2022}.

For NER, the mainstream ``fine-tuning''-based methods treat it as a sequence tagging problem \cite{NER_1,NER_2,NER_3}, while some approaches formalize it as a sequence-to-sequence task \cite{NER_4_seq2seq}.
Based on the ``prompt-learning'' paradigm, Chen et al. proposed a generative framework with learnable prompts \cite{lightNER} and achieved encouraging results on low-resource NER tasks. 

Same as NER, the vast majority of PLM-based RE methods use the ``fine-tuning'' paradigm \cite{re_fine_tuning_1,re_fine_tuning_2,re_fine_tuning_3}. 
In 2021, Han et al. proposed a text classification framework named ``PTR'' \cite{PTR_2021}, which could extract relations based on the ``prompt-learning'' paradigm.
A similar work following is ``Konwprompt'' \cite{knowprompt_2022}, which incorporates knowledge among relation labels into the prompt-learning paradigm.

Event Extraction is a more complex task due to the challenge of extracting multiple components.
According to the number of steps required to extract events, current approaches can be divided into two categories in structure.
The methods with pipeline structure obtain triggers first and then extract arguments based on triggers predicted \cite{DMCNN_Chen_2015,DEEB-RNN_Zhao_2018,PLMEE_Yang_2019}.
Another line of work \cite{JRNN_Nguyen_2016,JMEE_Liu_2018} with joint architecture extracts triggers and arguments simultaneously and is free from error propagation.
When it comes to the technology of extracting event components, conventional methods \cite{PLMEE_Yang_2019} obtain triggers and arguments by classifying tokens into specific categories.
Some works reformulate EE as a question answering \cite{QAEE_DU_20} or machine reading comprehension task \cite{MRCEE_Liu_2021}.
With the appearance of encoder-to-decoder-based PLMs such as T5 and BART, generative EE models \cite{Text2Event_2021, TANL_Paolini_2021} have been explored in recent years.
Finally, based on prompt learning paradigm, researchers convert EE to tasks that are familiar to PLMs and achieve encouraging results \cite{CondiGen_Li_2021, degree}.

UIE \cite{UIE_2022} is a unified generative framework for IE which extracts various forms of information by generating target structures.
To model different IE tasks universally, it designs a structural extraction language to encode heterogeneous information structures and a structural schema instructor to guide the generative model.
Although UIE is also a T5-based unified generative framework for IE, our CPGF is very different from UIE in the following respects:
\begin{itemize}
\item The requirements for training data are different. UIE pre-trains the backbone model on the large-scale corpus to learn common IE abilities from various knowledge sources. CPGF aims to obtain remarkable performance by training on the target dataset directly.
\item The unified forms are distinct. CPGF transforms IE into the pre-training task (slot-filling) of T5, while UIE outputs structural information via constraint generation. Therefore, our CPGF can elicit knowledge contained in the original T5 model more efficiently. 

\end{itemize}

% \begin{itemize}
% \item UIE and CPGF are designed with different intentions. UIE aims at learning general IE abilities from different knowledge sources. In contrast, CPGF is committed to developing a universal framework that can handle each IE task independently.
% \item The unified forms are different. CPGF transforms IE into the pre-training task (slot-filling) of T5, while UIE is constraint generation. Strictly speaking, UIE is not a framework based on the ``prompt learning'' paradigm.
% \item The training strategies are different. UIE relies on pre-training on the large-scale corpus to learn common IE abilities. CPGF is trained on the target dataset to obtain remarkable performance.
% \end{itemize}

% It is worth noting that Lin et al. proposed a prompt-based event extraction model named ``PoKE'' \cite{prompt_t5_ee}, which uses T5 as the PLM as well.
% It seems that the joint argument prompt of PoKE is similar to prompts we designed for Event Extraction. 
% However, there are actually many differences between us.
% Firstly, we work on designing prompts that can be used to extract events jointly, while their prompts are for Event Extraction models with the pipeline structure.
% Additionally, the prompt templates we used are type-specific and consist of multiple sub-prompts, which can further leverage the modular sub-prompt library to build composable prompts for data-scarce scenarios.

\section{Background}

\subsection{Prompt Learning Paradigm}

The main idea of the ``prompt learning'' paradigm is to translate downstream tasks into pre-training tasks (usually cloze) familiar to PLMs \cite{prompt_liupengfei}.
Take the widely used pre-training task ``Masked Language Model (MLM)'' as an example. The original text ${S}$ is converted to the input of a PLM by a pre-designed prompting function firstly:
$$S_{\text{prompt}} = f_{\text{prompt}}(S).$$
And then, the PLM predicts each masked word as it did in the pre-training process:
$$z=\text{MLM}(S_{\text{prompt}}),$$
where $\text{MLM}(\cdot)$ indicates the algorithm used for the MLM task during pre-training. 
A slight difference is that a set of permissible values $\mathcal{Z} = \{z_1, z_2, \dots \}$ is defined to constrain predictions when applying this paradigm for downstream tasks.
The values set $\mathcal{Z}$ could be a set consisting of many words in the case of classification, while the whole vocabulary when dealing with generative tasks \cite{prompt_liupengfei}.
Finally, a mapping function is required to obtain the label in some cases where the answer $\textit{z}$ is not the direct expected output.

% \begin{figure}[thb]

% \centering
% \includegraphics[width=1\columnwidth]{figs/task_examples.pdf} 
% %\vspace{-1em}% Reduce the figure size so that it is slightly narrower than the column. Don't use precise values for figure width.This setup will avoid overfull boxes.
% % \vspace{-2em}
% \caption{An example of extracting entity, event, relation from sentence.}
% \label{task_examples}
% \end{figure}

\subsection{Information Extraction}

Information Extraction is a text processing task, the purpose of which is to automatically extract specific types of facts from natural language texts and output structured results.
Facts come in many forms, such as entities, relationships, and events.
Based on the above three forms, Information Extraction consists of three essential subtasks: Named Entity Recognition, Relation Extraction, and Event Extraction.

\subsubsection{Named Entity Recognition}

The purpose of Named Entity Recognition is to automatically find the exact representation (words or phrases with specific meanings, such as a person, location, and organization\cite{CONLL2003}) of each entity from the text and determine its type \cite{ner_survey_2022}.
As shown in Fig.\ref{difference_from_current_methods}(a), an automated system for NER should recognize that "Malaysia" is an entity from the given text and classify it as "location."

\subsubsection{Event Extraction}

Event Extraction in this paper refers to automatically extracting structured event information from unstructured natural language texts under the guidance of an event schema.
Events information is more complex than entities.
% Events are more complex information than entities.
It is defined as a specific occurrence involving participants\footnote{http://projects.ldc.upenn.edu/ace/} and consists of a trigger and a set of arguments.
To make the Event Extraction task easier to understand, we introduce essential terminologies as follows:
\begin{itemize}
    \item \textbf{Trigger} indicates the word or phrase that best represents the occurrence of an event. 
    \item \textbf{Event type} is a category defined in the event schema. It is usually obtained by classifying a trigger. 
    \item \textbf{Event argument} is an entity participating in the event or an attribute value (such as TIME, CRIME, MONEY) of the event.
    \item \textbf{Argument role} refers to the role that an argument plays in the event it appears.   
\end{itemize}

Take Fig.\ref{difference_from_current_methods}(a) as an example. Extracting the event from this sentence consists of four items: 
\begin{itemize}
    \item Recognizing ``protests'' as the trigger.
    \item Classifying this event as type ``demonstrate''.
    \item Identifying that ``Retired military officers'' and ``Malaysia'' are arguments of this event.
    \item Assigning the roles ``agent'' and ``place'' to arguments ``Retired military officers'' and ``Malaysia''.  
\end{itemize}

\subsubsection{Relation Extraction}

Relation Extraction aims to assign a relation type to a pair of entities with logical connections \cite{RE_survey_2020}.
Given the organization entity ``Retired military officer'' and the country entity ``Malaysia'',  a Relation Extraction model should have the ability to judge that the relation between them is ``country\_of\_headquarters'' according to the context semantics.

\begin{figure*}[t]
\vspace{-1em}
\centering
\includegraphics[width=2\columnwidth]{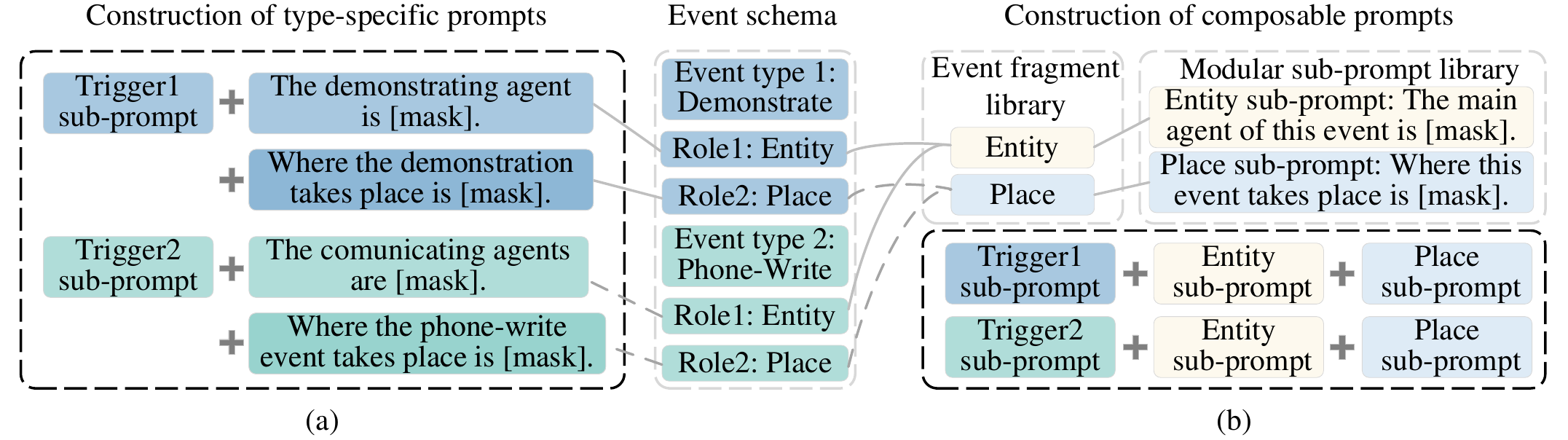} 
\vspace{-1em}
\caption{An example of constructing type-specific and composable prompts for Event Extraction.}
\label{prompt_compare}
\end{figure*}

\section{Unified Composable Prompt-learning-based Generative Framework}

% \begin{figure}[htb]
% % \vspace{-1em}
% \centering
% \includegraphics[width=1\columnwidth]{figs/overview_framework.pdf} 
% %\vspace{-1em}% Reduce the figure size so that it is slightly narrower than the column. Don't use precise values for figure width.This setup will avoid overfull boxes.
% % \vspace{-2em}
% \caption{The overview of our framework.}
% \label{overview_framework}
% \end{figure}

As shown in Fig.\ref{difference_from_current_methods}(e), the framework proposed in this paper consists of four components.
Information decomposition indicates splitting complex information into multiple fragments. 
Sub-prompts generation and Prompts Construction are the stages that yield sub-prompts for information fragments and type-specific or composable prompts (exclusive to EE) for information types, respectively.
In the slots filling stage, we apply the generative PLM T5 to predict answers.

\subsection{Task Formalization}\label{sec:task_form}

% This section will first introduce the information extraction framework proposed on a macro level.
% Then, a detailed description of the prompts and outputs construction will be presented, followed by a detailed description of Word Hint strategy and training.
% The last part is the application of PGF on the event extraction task.

% \subsection{Overview of PGF}

Given an origin text ${S}$, the purpose of Information Extraction is to obtain target information $\mathcal{Y}=\{\mathcal{Y}^1, \dots, \mathcal{Y}^{t}\}$, where $\mathcal{Y}^j, j\in [1,t] $ represents the information to extract for the $j$-th type, and $t$ refers to the number of types. 
We build a set of prompts ${Pr}=\{{Pr}^1, \dots, {Pr}^{t}\}$ to extract information by filling slots in the prompts.
According to the characteristics of specific tasks, we regard the target information as a collection of multiple minimum granularity fragments: $\mathcal{Y}^j=\{y_1^j, y_2^j, \dots, y_n^j\}$, where $n$ is the number of components that make up the information with the $j$-th type.
For Event Extraction, the components set for the event with type of ``$\text{event\_j}$'' could be represented as:
\begin{equation}
    \begin{aligned}
        \mathcal{Y}^{\text{event}\_j}=& \{\text{trigger}\mbox{-} \text{event~type}^j, \text{argument}\mbox{-} \\
                                  & \text{role}_1^j, \dots, \text{argument}\mbox{-} \text{role}_{m^j}^j\}, \nonumber
    \end{aligned}
\end{equation}
where $m^j$ indicates the number of arguments participating in the event type. For NER and RE, since an entity and relation cannot be decomposed, only one element is included in their component set.

\subsection{Sub-prompts Generation}\label{sec:sub-prompt-gen}

This subsection will introduce two kinds of sub-prompts for complex IE tasks such as Event Extraction.
The type-dependent sub-prompts are suitable for data-abundant scenarios, while modular sub-prompts are designed for data-scarce scenarios.
Fig.\ref{prompt_compare} shows the difference between these two kinds of sub-prompts.
The left part of this figure shows the relationship between type-dependent sub-prompts and information categories.
In this situation, sub-prompts are independently generated by replacing keywords in definitions or descriptions of fragments in each information type with mask words.

In the case of using modular sub-templates, we merge similar information fragments before yielding sub-prompts. 
As shown in Fig.\ref{prompt_compare}(b), CPGF treats elements with the same semantics but are distributed in different types of information as the same information fragment.
We manually build an information fragment library that covers all compositions of each information type by analyzing the semantics of the information elements in the training dataset.
Each type of information in the dataset is composed of one or more pieces of information fragments in the library.

We make up a modular sub-prompt library ${P}=\{p_1, p_2, \dots, p_{|F|}\}$, where $|F|$ indicates the total number of information fragments, by generating a type-independent sub-prompt for every element in the information fragment library.
In detail, it replaces the keywords in a generic sentence that describes the relationship between the fragment and the information types to get a modular sub-prompt.
Our framework yields sub-prompts for target information by searching its fragments from the modular sub-prompt library.
In this way, CPGF can efficiently extract the elements of information with types never seen if these elements appear in the fragment library.

\subsection{Prompts Construction}

% To incorporate more intuitive semantic information into the prompt templates and give our framework the ability to extract multiple types of information independently, PGF applies a prompting function set $\mathcal{F}$ to generate type-specific prompts.
% In detail, we firstly generate a sub-prompt $p_i^j$ for the $i$-th element in $\mathcal{Y}^j$ by its corresponding sub-prompting function $f_{\text{sub}\_\text{prompt}\_i}(\cdot)$.
% $$p_i^j=f_{\text{sub}\_\text{prompt}\_i}^{j}(y_i).$$

The prompt for the $j$-th information type is constructed by connecting the original text with sub-prompts obtained in subsection \ref{sec:sub-prompt-gen}:
% \vspace{-0.5em}
\begin{equation}
    \begin{aligned}
        \mathcal S_{\text{prompt}}^{j} &= f_{\text{prompt}}(S) \\
            &= S~\mathop{||}~Pr^j, \nonumber
    \end{aligned}
    % \vspace{-0.5em}
\end{equation}
where $\mathop{||}$ is defined as the operation of concatenating two texts, and $Pr^j$ represents the result of combining all the sub-prompts that related to elements contained in the target information:
\begin{equation}
    \begin{aligned}
         Pr^j =  \mathop{||}\limits_{i=1}\limits^{n} p_i^j, \nonumber
    \end{aligned}
\end{equation}
% The sub-prompts above are selected from the modular sub-prompt library when constructing a composable prompt.
where the value of $n$ is always $1$ for NER and RE.
For Event Extraction, we called $S_{\text{prompt}}^{j}$ ``type-specific prompt'' if sub-prompts used to consist it are type-dependent because each part of this prompt is related to the information type.
Prompts composed of modular sub-prompts are named ``Composable prompts''.

% Following the above steps, PGF proposed in this paper generates a prompt for each type.

% It is worth mentioning that the approach of constructing prompts we proposed in this paper have at least two advantages: 1) the of target types can be intuitively described in sub-prompts to enhance the suggestiveness of prompts.
% 2) the type-specific prompts

\subsection{Answer Generation}

After obtaining a set of prompts for a sample, we feed them to a generative PLM.
The PLM outputs a sequence consisting of predicted values for masked words in each prompt.
Specifically, when processing the prompt for the $j$-th type, the decoder of the generative PLM constructs the output sequence by successively predicting slots in sub-prompts:
$$z_i^j=\text{MLM}(p_i^j).$$
In this progress, we apply the special character `$|$' to separate predicted results for the sub-prompt, which has multiple answers.
% To ensure the uniqueness of the output, we stipulate that answers in the output sequence must follow their corresponding mask tokens such as ``$[\text{mask}_0]~ \text{answer}_0~|~\text{answer}_1$''.

After that, our framework maps each output to a component of target information, which could consist of one or multiple values.
Finally, we manually aggregate components obtained into complete information $\mathcal{Y}^j$.
% Finally, we manually aggregate them into complete information $\mathcal{Y}=\{\mathcal{Y}^1, \dots, \mathcal{Y}^{|\text{types}|}\},$ where $|\text{types}|$ represents the number of types, and $\mathcal{Y}^j=\{y_1^j, \dots, y_n^j\}$ indicates the information extracted for the $j$-th type.

% For example, when extracting relationships of entities, we could build many specific prompts, of which the prompt for the relation type ``leader'' is ``It is [MASK] that Joe Biden is the \textit{leader} of America.''.
% In this case, the only element in the information component set is ``relationship''.
% Machines can infer the existence of the relationship of ``leader'' if the prediction of this prompt is "correct".

\begin{figure*}[thb]
\vspace{-1em}
\centering
\includegraphics[width=2\columnwidth]{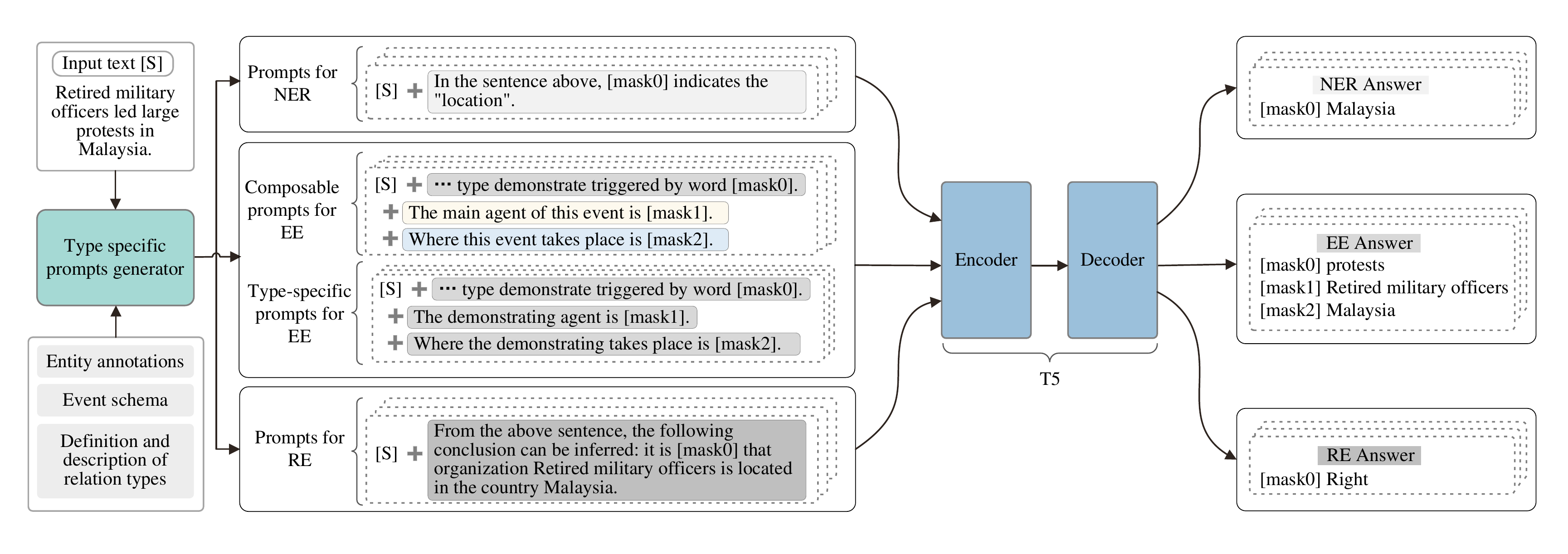} 
%\vspace{-1em}% Reduce the figure size so that it is slightly narrower than the column. Don't use precise values for figure width.This setup will avoid overfull boxes.
\vspace{-1em}
\caption{Schematic diagram of extracting entities, events, and relationships using CPGF. The complete sub-prompt for the trigger of the ``demonstrate'' event is ``There is an event with type demonstrate triggered by the word [mask0]''.   }
\label{overviewer}
\end{figure*}

\subsection{Training}

The ultimate purpose of the generative PLM in our framework is to output a token sequence consisting of information fragments and mask tokens that equal the golden one.
To fully use the dependence among the answers of sub-prompts, slots-filling is formalized as an autoregressive text generation in our framework.
% Take Event Extraction as an example. 
Inputting the prompt for $\textit{t}$-th information type $S_{\text{prompt}}^t$, the conditional probability of the output sequence is progressively combined by the probability of each token:
\begin{equation}
   p(Q|S_{\text{prompt}}^t)= \prod_{i=1}^{|Q|} p(q_i | q_{<i}, S_{\text{prompt}}^t), \nonumber
\end{equation}
where $Q$ is the sequence to generate, $q_i$ is the $i$-th token in sequence $Q$, $q_{<i}$ represents the tokens in the front of $q_i$ and $p(q_i | q_{<i}, S_{\text{prompt}}^t)$ is the conditional probability for token $q_i$.

In the stage of training, the decoder predicts each word with golden tokens before the current position.
During inference, the sequence is generated by applying greedy decoding.
It means that the decoder selects a token with the highest probability from the vocabulary of the generative PLM as the output at the current moment according to the subsequence generated.

The training object is to minimize the loss function, which can be formalized as:
\begin{equation}
   \mathcal{L} = -\sum_{i=1}^{|D|} \sum_{j=1}^{|Q_i|} \sum_{k=1}^{|V|} q_{i,j,k} \log p(q_{i,j,k} | q_{i,<j}, S_{\text{prompt}}^t), \nonumber
\end{equation}
where $|D|$, $|Q_i|$, and $|V|$ are the symbolic representations of the size of the training set, the length of the $i$-th output sequence, and the number of tokens in the vocabulary, respectively.

% \subsection{Word Hint Strategy (need modify again)}

% We conduct inference on the training set after completing the training for an epoch.
% The Word Hint strategy will be triggered whenever there are any errors in a predicted sequence.
% The uncertain samples in which the model has made mistakes will be regenerated in this stage. 
% In detail, for the case that the correct answer for the uncertain sub-prompt has multiple words, the Word Hint strategy inserts the first word of the golden answer in front of the mask word in the sub-prompt.
% On the one hand, this strategy reduces the difficulty of sub-prompts for uncertain samples..
% On the other hand, it increases the variety of sub-prompts for uncertain samples.
% % The presence of this hint word brings reference information, which can be used in generating the answer for the uncertain sub-prompt during the next epoch, to the decoder of the generative PLM.
% % It prevents the decoder from using predicted subsequences, which have the probability of containing errors, as the only reference information when predicting new words.

% When dealing with the uncertain sub-prompt whose answer has only one word, we replace the mask word with the golden answer.
% In this situation, the decoder will not generate an answer for the sub-prompt.
% This practice ensures that, in the next epoch, the possible incorrect answer of this uncertain sub-prompt will no longer affect the answer generation of other sub-prompts.

\subsection{Applying on Information Extraction tasks}

\subsubsection{Event Extraction}

Given a sentence consisting of token sequence $S=\{w_1,w_2,\ldots,w_{|S|}\}$, the EE task aims to obtain events $\mathcal{Y}=\{e_1,e_2,\ldots,e_{|E|}\}$, where each event can be written as $e_i=\{tr^i,Ar^i\}$.
In more detail, $tr^i$ is the trigger of the $i$-th event, and the set of arguments $Ar^i$ can be further expressed as $Ar^i=\{ar^i_1,ar^i_2,\ldots,ar^i_{|Ar^i|}\}$.
Furthermore, $tr^i$ and $ar^i_j$ are sub-sequences of sentence $S$, and approaches need to map them to the correct event type and argument role.

% The generative PLM used in applying examples in this paper is T5.
Fig.\ref{overviewer} shows the process of extracting events with composable and type-specific prompts. 
Given a text $S$, We obtain the sub-prompt for a trigger by replacing the keywords in a declarative sentence that describes the event's occurrence with a mask token ``$\left \langle extra\_id\_i \right \rangle$'' defined by T5.
There are two kinds of sub-prompts for argument roles: modular and type-dependent.
We first build the modular sub-prompts library for argument roles of all event types in the event schema if using composable prompts.
The stems of modular sub-prompts for argument roles are universal annotation guidelines that elaborate on the roles' implications in related events.
For type-dependent sub-prompts, we use more detailed descriptions (including event types' names) as stems for elements in each event.
To design the question for the cloze task, we replace the critical positions in the above two kinks of stems with mask tokens.
Then we generate composable or type-specific prompts for each event type by connecting the original text, tigger sub-prompt, and corresponding modular or type-dependent sub-prompts orderly.
Next, the generative model T5 gets the concatenated texts as inputs and outputs the sequences consisting of answers for sub-prompts.
% We apply the special character `$|$' to separate the multiple triggers that indicate events with the same type or arguments that play the same role in an event.
% To ensure the uniqueness of the output, we stipulate that answers in the output sequence must follow their corresponding mask tokens.

By analyzing output sequences, our framework can obtain fragments about triggers, event types, event arguments, and argument roles.
For instance, the predicted answer ``Malaysia'' for the sub-prompt ``Where this event takes place $\left \langle extra\_id\_2 \right \rangle$'' indicates that ``Malaysia'' plays a role of ``place'' in the ``demonstrate'' event. 

% \begin{figure}[thb]
% \centering
% \includegraphics[width=1\columnwidth]{figs/ner_re_overview.pdf} 
% %\vspace{-1em}% Reduce the figure size so that it is slightly narrower than the column. Don't use precise values for figure width.This setup will avoid overfull boxes.
% \caption{The overview of applying CPGF on the NER and RE task.}
% \label{ner_re_overview}
% \end{figure}

\subsubsection{Named Entity Recognition}

Since an entity cannot be split into multiple fragments, our framework treats the NER task as extracting an argument or a trigger from an input sentence.
As shown in Fig.\ref{overviewer}, the application of CPGF to the NER task is very similar to that of EE, except that each prompt for NER has only one sub-prompt.
We consider ``In the sentence above, words $\left \langle extra\_id\_0 \right \rangle$ indicate the (entity type).'' as the prompt template in this paper.

\begin{figure*}[thb]
\vspace{-1em}
\centering
\includegraphics[width=2\columnwidth]{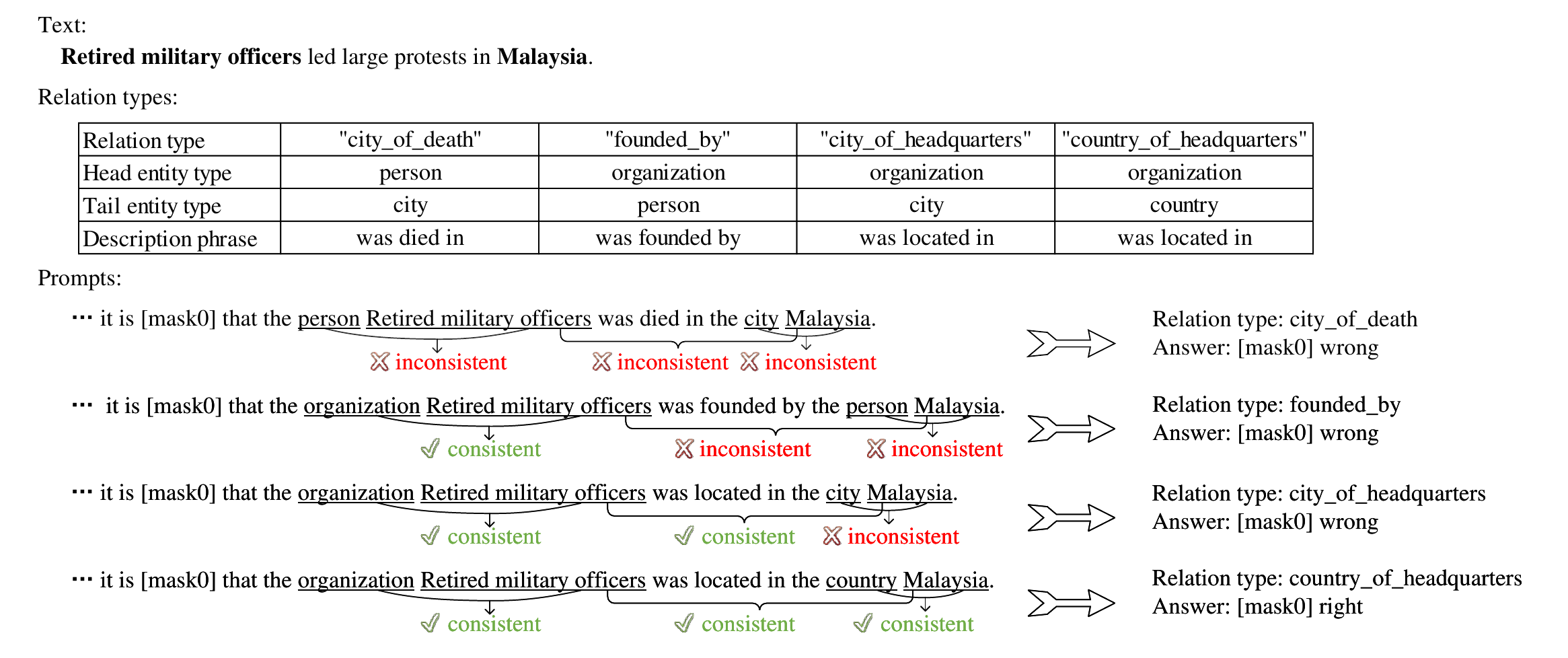} 
%\vspace{-1em}% Reduce the figure size so that it is slightly narrower than the column. Don't use precise values for figure width.This setup will avoid overfull boxes.
\vspace{-1em}
\caption{An example of judging the existence of relation types from inconsistencies in the prompts.  }
\label{re_prompt_example}
\end{figure*}

\subsubsection{Relation Extraction}

Consider that Relation Extraction aims to classify relation types between entity pairs rather than extract spans from given texts.
It is challenging for generative models to predict relation types' names directly.
To solve this problem, we design a sophisticated template that allows our framework to classify relationships by predicting whether contextual semantics are conflicting.

Fig.\ref{re_prompt_example} shows the details of the prompt template and illustrates the logic of extracting relations by estimating semantic inconsistencies in the prompts.
The types of head and tail entities are known knowledge for a relationship.
We first concatenate these types with mentions of head and tail entities in the text, respectively.
Then, we connect the new representations of these two entities with a phrase representing the relation type's semantics.
The resulting text is expanded into a subject clause containing a ``[mask]'' token.

We can see from the example in Fig.\ref{re_prompt_example} that since there is no relation ``founded\_by'' between the entity pairs, the prompt constructed by the above approach will be semantically inconsistent with the original text.
First, entity type ``person'' and entity mention ``Malaysia'' are inconsistent.
Furthermore, the semantics of the connecting phrase ``was
founded by'' conflicts with the relationship implied in the original text.
In this case, the golden answer for ``[mask0]'' is ``wrong''.
The golden answer is ``right'' only if there are no inconsistencies in the prompt.

% Fig.\ref{re_prompt_example} shows a detailed example of extracting relation by applying our CPGF.
% We first connect two given entities with a phrase ``is located in'' that expresses the semantics of the relation type ``country of headquarters''. 
% The resulting text is then expanded into a subject clause containing a ``[MASK]'' token.
% Our framework determines whether the original input ``[S]'' is semantically conflicting with the prompt by predicting the ''[MASK]'' token as ``right'' or ``wrong''.
% If the output is ``right'', then the semantics of the original input are consistent with the that of the prompt.
% That is, there is a relationship of the current type between the entity pair.
% Rather the opposite.

\section{Experiments}

\subsection{Experimental Setups}\label{sec:exp_setup}

% \subsubsection{Dataset and evaluation metric}

We evaluated the performance for NER on the dataset CoNLL2003 with the standard split the same as LightNER.
Experiments for Event Extraction were conducted on the widely used dataset ACE2005, which covers 7 types and 33 subtypes.
For the RE task, we selected the traditional corpus SemEval \cite{SemEval_2010} as the dataset. It contains 9 bidirectional relation types and a special relation ``Other''.
To make a fair comparison, we applied the same split as the previous works \cite{oneie_Lin_2020, PTR_2021, QAEE_DU_20, Text2Event_2021, degree} and followed the evaluation criteria used in them as well.

\begin{table}[t]
\renewcommand\arraystretch{1.2}
\centering
\caption{Hyper-parameter searching spaces for IE tasks. }
\label{parameters_table}
\setlength{\tabcolsep}{3mm}{
\begin{tabular}{llc}
\toprule

Parameter              & Task                & Values \\ \midrule
\multicolumn{3}{c}{Data abundant} \\ \midrule
Learning rate & NER, EE, RE             & [5e-5,1e-4]  \\ \hline
Weight decay  & NER, EE, RE              & [1e-4,1e-3,1e-2]  \\ \hline 
%\cline{2-3} 
Batch size for T5 base  & NER, EE, RE              & [32,64]  \\ 
Batch size for T5 large   & NER, EE, RE                  & [20] \\ \midrule

\multicolumn{3}{c}{Low-resource} \\ \midrule
\multirow{2}{*}{Learning rate} & NER, EE & [1e-4, 3e-4, 5e-4] \\ 
& RE & [3e-5, 5e-5, 7e-5, 1e-4] \\ \hline
\multirow{2}{*}{Weight decay}  & NER, EE               & [1e-3, 1e-2, 1e-1]  \\ 
%\cline{2-3} 
 & RE & [3e-2, 1e-1, 3e-1]  \\ \hline
\multirow{2}{*}{Batch size}  & NER, EE              & [4,8] \\ 
 & RE  & [32] \\ 
\bottomrule
\end{tabular}}
\vspace{-1em}
\end{table}

We utilized a single NVIDIA RTX A6000 for the training/evaluation on the full dataset and a single NVIDIA TITAN RTX for the experiments on data-scarce scenarios. 
The optimizer in experiments is AdamW. We used the pre-trained parameters ``T5-base'' and "T5-large" to initialize the T5 model.
In this paper, ``CPGF-base'' stands for using ``T5-base'' as the PLM in the CPGF framework with type-specific prompts, and ``CPFG-large'' represents applying ``T5-large''.
In all scenarios of IE tasks, the optimal hyperparameters are obtained by grid searching with the F1-score as the criterion. The search spaces are shown in Table.\ref{parameters_table}. 
% Table~\ref{parameters_table} shows the hyperparameter settings in all scenarios.

\subsection{Results for Event Extraction}

\subsubsection{Baselines}

To verify the advantages of CPGF proposed in this paper, we selected methods that embody generative, prompt learning, and fusing label information as baselines.
We compared CPGF-based Event Extraction models with four similar prior models: EEQA, OneIE, Text2Event, and DEGREE in data-abundant and data-scarce scenarios. 
EEQA \cite{QAEE_DU_20} formulates Event Extraction as a question-answering task, and OneIE is a system trained on multi-task with global features.
Text2Event \cite{Text2Event_2021} is a generative method that converts Event Extraction into a sequence-to-structure problem and outputs structured event information by a constrained decoding algorithm.
``Text2Event-B'' and ``Text2Event-L'' indicate two results obtained by using ``T5-base'' and ``T5-large''.
DEGREE \cite{degree} is the state-of-the-art prompt-learning-based EE method, but it does not cover all the argument roles of the dataset ACE2005. 
For fairness, we obtained the results of DEGREE by rerunning its code but evaluating on all argument roles used in the other three baselines.
DEGREE-a indicates the results of retraining DEGREE with new templates that include missing roles of the original template. 
Since UIE is pre-trained for IE on a large-scale corpus, comparing it with other works selected in this paper is unfair. 
Therefore we do not consider it as a baseline in this paper.

\begin{table}[t]
\vspace{-1em}
\renewcommand\arraystretch{1}
\centering
\caption{EE results(\%) in the data-abundant scenario on ACE2005.}
\label{supervised_result_table}

\resizebox{1\columnwidth}{!}{
\begin{tabular}{lcccccc}
\toprule
\multirow{2}{*}{Model} & \multicolumn{3}{c}{T-C}                                      & \multicolumn{3}{c}{A-C}                                      \\ \cline{2-7} 
                       & P    & R    & F1    & P    & R    & F1   \\ \hline
OneIE                  & \multicolumn{1}{c}{-}    & \multicolumn{1}{c}{-}    & 74.70  & \multicolumn{1}{c}{-}    & \multicolumn{1}{c}{-}    & \textbf{56.80}\\
Text2Event-B             & \multicolumn{1}{c}{67.50} & \multicolumn{1}{c}{71.20} & 69.20  & \multicolumn{1}{c}{46.70} & \multicolumn{1}{c}{53.40} & 49.80 \\ 
Text2Event-L             & \multicolumn{1}{c}{69.60} & \multicolumn{1}{c}{74.70} & 71.90  & \multicolumn{1}{c}{52.50} & \multicolumn{1}{c}{55.20} & 53.80 \\ 
EEQA                   & \multicolumn{1}{c}{71.12} & \multicolumn{1}{c}{73.70} & 72.39 & \multicolumn{1}{c}{56.77} & \multicolumn{1}{c}{50.24} & 53.31 \\ 
% CondiGen              & \multicolumn{1}{c}{-}    & \multicolumn{1}{c}{-}    & 71.1  & \multicolumn{1}{c}{-}    & \multicolumn{1}{c}{-}    & 53.7 \\ 
% TANL                   & \multicolumn{1}{c}{-}    & \multicolumn{1}{c}{-}    & 68.5  & \multicolumn{1}{c}{-}    & \multicolumn{1}{c}{-}    & 48.5 \\
DEGREE$^{\dagger}$      & \multicolumn{1}{c}{76.19}    & \multicolumn{1}{c}{68.37}    & 72.07  & \multicolumn{1}{c}{53.23}    & \multicolumn{1}{c}{46.24}    & 49.49 \\
DEGREE-a$^{\dagger}$   & \multicolumn{1}{c}{76.43}    & \multicolumn{1}{c}{71.36}    & 73.81  & \multicolumn{1}{c}{58.19}    & \multicolumn{1}{c}{48.37}    & 52.82 \\\hline
CPGF-base  & \multicolumn{1}{c}{75.34} & \multicolumn{1}{c}{79.62} & 77.42 & \multicolumn{1}{c}{53.45} & \multicolumn{1}{c}{53.51} & 53.48 \\
% ~~~ \textit{w/} hint    & \multicolumn{1}{c}{77.36} & \multicolumn{1}{c}{77.73} & 77.54 & \multicolumn{1}{c}{56.06} & \multicolumn{1}{c}{52.66} & 54.31 \\
% ~  \textit{w/} modular    & \multicolumn{1}{c}{0} & \multicolumn{1}{c}{0} & 0 & \multicolumn{1}{c}{0} & \multicolumn{1}{c}{0} & 51.5 \\
CPGF-large  & \multicolumn{1}{c}{80.81} & \multicolumn{1}{c}{75.83} & \textbf{78.24} & \multicolumn{1}{c}{57.69} & \multicolumn{1}{c}{51.33} & 54.32 \\
% ~~~  \textit{w/} hint    & \multicolumn{1}{c}{0} & \multicolumn{1}{c}{0} & 0 & \multicolumn{1}{c}{0} & \multicolumn{1}{c}{0} & 0 \\
% ~  \textit{w/} modular    & \multicolumn{1}{c}{0} & \multicolumn{1}{c}{0} & 0 & \multicolumn{1}{c}{0} & \multicolumn{1}{c}{0} & 51.9 \\
\bottomrule
\end{tabular}
}
% }
\vspace{-1em}
\end{table}

\subsubsection{Data Abundant}

Table~\ref{supervised_result_table} shows the overall performance on the complete ACE2005 dataset.
T-C and A-C stand for trigger classification and argument classification, respectively.
We can observe that CPGF-base and CPGF-large significantly outperform the best F1 in baselines for trigger classification by 2.72\% and 3.54\%, respectively.
Although both DEGREE and CPGF are prompt-based methods, our CPGF-large outperforms DEGREE by 4.43\% in trigger classification.
Considering that more than 99.5\% of the triggers in the ACE2005 corpus consist of only one word.
We believe the reason for the substantial improvement is that applying T5 to complete the cloze task is more conducive to extracting triggers than using BART to recover templates.

CPGF-based models achieve excellent F1 scores for argument extraction as well.
% Specifically, compared to Text2Event, which is also based on T5, CPGF performs better when applying the same PLM. 
On the same training data, our models considerably improve the F1 score of argument classification when compared with DEGREE.
Notice that the model OneIE is trained on multi-tasks with entity, relation, and event annotations.
Therefore, as far as the Event Extraction task is concerned, the disadvantage of our method in the F1 score does not mean that our approach is inferior to OneIE in performance on argument extraction.
The results above illustrate the effectiveness of our framework.

\begin{table}[t]
\vspace{-1em}
\renewcommand\arraystretch{1}
\caption{Results(\%) of zero shot Event Extraction.}
\label{zero_shot_result_table}
\resizebox{1\columnwidth}{!}{
\begin{tabular}{lcccccc}
\toprule
\multirow{2}{*}{Model} & \multicolumn{3}{c}{T-C}                                      & \multicolumn{3}{c}{A-C}                                      \\ \cline{2-7} 
                       & P    & R    & F1    & P    & R    & F1   \\ \hline
Transfer                  & \multicolumn{1}{c}{75.50}    & \multicolumn{1}{c}{36.30}    & 49.10  & \multicolumn{1}{c}{16.10}    & \multicolumn{1}{c}{15.60}    & {15.80}\\
ILP             & \multicolumn{1}{c}{54.10} & \multicolumn{1}{c}{53.10} & 53.60  & \multicolumn{1}{c}{4.60} & \multicolumn{1}{c}{10.00} & 6.30 \\ 
TE/QA                   & \multicolumn{1}{c}{-} & \multicolumn{1}{c}{-} & 41.70 & \multicolumn{1}{c}{-} & \multicolumn{1}{c}{-} & 16.80 \\ 
DEGREE$^{\dagger}$      & \multicolumn{1}{c}{52.47}    & \multicolumn{1}{c}{53.75}    & 53.11  & \multicolumn{1}{c}{45.10}    & \multicolumn{1}{c}{15.34}    & 22.89 \\
DEGREE-a$^{\dagger}$   & \multicolumn{1}{c}{62.13}    & \multicolumn{1}{c}{44.97}    & 52.17  & \multicolumn{1}{c}{45.19}    & \multicolumn{1}{c}{16.70}    & 24.39 \\\hline
CPGF-base  & \multicolumn{1}{c}{47.86} & \multicolumn{1}{c}{52.30} & 49.98 & \multicolumn{1}{c}{32.38} & \multicolumn{1}{c}{27.87} & 29.95 \\
~  \textit{w/} composable    & \multicolumn{1}{c}{47.48} & \multicolumn{1}{c}{51.59} & 49.45 & \multicolumn{1}{c}{33.44} & \multicolumn{1}{c}{30.46} & 31.88 \\
% base \textit{w/} hint    & \multicolumn{1}{c}{55.8} & \multicolumn{1}{c}{44.6} & 49.6 & \multicolumn{1}{c}{34.1} & \multicolumn{1}{c}{22.4} & 27.0 \\
CPGF-large  & \multicolumn{1}{c}{56.14} & \multicolumn{1}{c}{55.87} & \textbf{56.01} & \multicolumn{1}{c}{38.37} & \multicolumn{1}{c}{32.41} & 35.14 \\
~  \textit{w/} composable    & \multicolumn{1}{c}{56.42} & \multicolumn{1}{c}{54.37} & 55.37 & \multicolumn{1}{c}{36.19} & \multicolumn{1}{c}{39.07} & \textbf{37.58} \\
% large \textit{w/} hint    & \multicolumn{1}{c}{51.4} & \multicolumn{1}{c}{52.1} & 51.8 & \multicolumn{1}{c}{37.0} & \multicolumn{1}{c}{34.6} & 35.7 \\
\bottomrule
\end{tabular}}
\vspace{-1em}
\end{table}

\subsubsection{Zero Shot}

% Zero-shot learning is one of the most challenging tasks in artificial intelligence. 
In this scenario, models are trained on many given types but evaluated on unseen types. 
We follow the data split used in paper \cite{Transfer_Huang_2018}: the top 10 most frequent event types are used during training, and the rest make up the test set. 
Three recent pieces of studies, Transfer \cite{Transfer_Huang_2018}, ILP \cite{ILP_zhang_2021}, and TE/QA \cite{TE/QA_Lyu_2020}, which are devoted to EE in the zero-shot scenario, are considered as baselines along with DEGREE. 
``\textit{w/} composable'' represents using composable prompts instead of type-specific prompts.

% To adapt to this scenario, we make several slight adjustments to the model.
% In terms of model structure, the "Word Hint" module is abandoned to improve the model's generalisation performance.
%We removed the correct examples in the template because in this scenario the examples of the types in the test set cannot be obtained.
As shown in Table~\ref{zero_shot_result_table}, both CPGF-base and CPGF-large achieve good F1 scores in T-C and A-C.
It is worth noting that, with type-specific prompts, our CPGF-based models have significant improvements in argument extraction compared to DEGREE.
It demonstrates that the method extracting arguments by multiple independent sub-prompts has better transferability than DEGREE.
Furthermore, as we can see intuitively from the table, the application of composable prompts significantly improves the F1 score of argument classification on both CPGF-base and CPGF-large models.
These increases occur because the modular sub-prompts cover some argument roles in the unseen event types.
The improvements prove the effectiveness of the composable prompts we proposed in the zero-shot setting.
We observe that the F1 values of T-C decrease slightly with the application of composable prompts.
The reasons for this phenomenon are that (1) event type is the unique element of an event and cannot benefit from modular sub-prompts, and (2) the application of modular sub-prompts results in a loss of semantics associated with event types.
However, relative to the improvement on the A-C, the slight decreases in triggers classification are acceptable.

\begin{table}[t]
\vspace{-1em}
\renewcommand\arraystretch{1}
\centering
\caption{F1 scores(\%) of Event Extraction in the scenario of low resource. The second best scores are underlined.}
\label{low_resource_result_table}
\setlength{\tabcolsep}{1mm}{
\begin{tabular}{l|cc|cc|cc|cc}
% \hline
\toprule
\multirow{2}{*}{Model} & \multicolumn{2}{c|}{1\%}        & \multicolumn{2}{c|}{5\%}         & \multicolumn{2}{c|}{10\%}        & \multicolumn{2}{c}{20\%}        \\
                       & T-C  & A-C & T-C  & A-C  & T-C  & A-C    & T-C  & A-C \\ \midrule
OneIE$^{\dagger}$                  & 38.16 & 7.16 & 57.26 & 22.96 & 58.14 & 27.16 & 66.58 & 33.91 \\
Text2Event-B$^{\dagger}$             & 14.60 & 6.20 & 43.78 & 20.45 & 46.14 & 23.68 & 54.46 & 34.24   \\ 
Text2Event-L$^{\dagger}$             & 20.96 & 9.66 & 46.66 & 23.64 & 48.07 & 26.79 & 57.24 & 35.35   \\ 
EEQA$^{\dagger}$                   & 18.21 & 6.33 & 52.91 & 18.78 & 54.09 & 20.49 & 61.27 & 24.27    \\ 
DEGREE$^{\dagger}$            & 50.74 & 15.84 & 63.30 & 33.45 & 65.07 & 37.20 & 66.53 & 41.52    \\
DEGREE-a$^{\dagger}$        & \underline{51.52} & 18.9 & \underline{64.41} & \textbf{35.70} & 65.61 & 38.46  & 66.92  & \underline{42.43}    \\  \midrule
CPGF-base    & 50.54 & 17.05   & 64.22 & 32.53 & 67.67 & 37.24    & 69.29 & 39.95    \\ 
% ~ \textit{w/} hint  & 55 & 17.4 & 62.7 & 32.6  & 67.7 & 34.7 & 70.5 & 40.2 \\ 
~ \textit{w/} composable  & \textbf{51.80} & 18.74 & \textbf{66.52} & 33.74  & 67.51 & \underline{38.68} & \textbf{71.36} & \textbf{43.30} \\
CPGF-large & 50.04 & \textbf{20.84}  & 63.00 & 33.24  & \textbf{69.58} & 37.69    & \underline{71.17} & 41.73    \\ 
~ \textit{w/} composable  & 48.13 & \underline{19.33} & \underline{64.41} & \underline{34.39}  & \underline{69.26} & \textbf{39.91} & 70.92 & 42.36 \\
% ~ \textit{w/} hint  & 47.6 & 21.2 & 66.8 & 32 & 69.2 & 40.1  & 71.4 & 41.7 \\
\bottomrule
\end{tabular}
}
\vspace{-1em}
\end{table}

\subsubsection{Low Resource}

% In the scenario of low resource, the challenge is to train an effective model on a tiny minority of data.
We conducted experiments with the same setup used in DEGREE \cite{degree} to build the training set.
In detail, we used 1\%, 5\%, 10\%, and 20\% of the training set as supervisory information in four experiments and kept the test and development set unchanged.
The random selection of training data has a great impact on performance.
Therefore, we retrained the previous methods with the same training data and evaluate them with the same criteria.
% Due to the lack of the data preprocessing module and the trigger extraction module in the public code, we cannot obtain the performances of TANL\footnote{https://github.com/amazon-research/tanl} and CondiGen \footnote{https://github.com/raspberryice/gen-arg} in low-resource scenarios.

We can observe from Table~\ref{low_resource_result_table} that our models with type-specific prompts achieve competitive results when compared to DEGREE.
% Compared to other baselines, CPGF-based models improves F1 scores of both triggers and arguments classification significantly regardless of the experimental setting.
This proves that our framework has a similar ability to adapt to low-resource scenarios as DEGREE.
When utilizing T5-large as the PLM, the model performance of T-C in 1\% and 5\% scenarios decreases slightly.
The reason is that T5-large suffers from overfitting due to the scarcity of trigger samples.
This problem is alleviated with increasing the size of training data.
The application of composable prompts brings noticeable improvements for argument extraction in almost all situations.
The only exception is the CPGF-large model in the 1\% data scenario.
We believe that in the case of extremely sparse data, modular sub-prompts reduce the diversity of samples, thereby exacerbating the degree of overfitting of the T5-large.
This observation demonstrates that constructing prompts with modular sub-prompts can enhance models' generalization ability.

\subsection{Results for Named Entity Recognition}

\subsubsection{Baselines}
For Named Entity Recognition, four strong methods, including LightNER \cite{lightNER}, were considered as baselines.
LC-BERT and LC-BART, used as baselines in LightNER, are the adoptions of label-specific classifiers on BERT and BART, respectively.
``Template'' \cite{template_BART} extracts entities by enumerating all possible spans and taking the span with the largest possibility as the final entity.

\begin{table}[thb]
\vspace{-1em}
\centering
\renewcommand\arraystretch{1}
\caption{Results (\%) for NER on the complete CoNLL2003 dataset.}
\label{ner_fully_result_table}
\setlength{\tabcolsep}{3.5mm}{
% \resizebox{1\columnwidth}{!}{

\begin{tabular}{lccc}
\toprule
Model   & P & R & F1 \\ \hline
LC-BERT                  & \multicolumn{1}{c}{91.93}    & \multicolumn{1}{c}{91.54} & 91.73\\
LC-BART             & \multicolumn{1}{c}{89.60} & \multicolumn{1}{c}{91.63} & 90.60\\ 
Template                   & \multicolumn{1}{c}{90.51} & \multicolumn{1}{c}{93.34} & 91.90 \\ 
LightNER      & \multicolumn{1}{c}{92.39}    & \multicolumn{1}{c}{\textbf{93.48}}    & \textbf{92.93} \\ \hline
CPGF-base  & \multicolumn{1}{c}{92.24} & \multicolumn{1}{c}{92.37} & 92.31 \\
% base \textit{w/} hint    & \multicolumn{1}{c}{0} & \multicolumn{1}{c}{0} & 0  \\
CPGF-large  & \multicolumn{1}{c}{\textbf{92.51}} & \multicolumn{1}{c}{93.35} & \textbf{92.93} \\
% large \textit{w/} hint    & \multicolumn{1}{c}{0} & \multicolumn{1}{c}{0} & 0 \\
\bottomrule
\end{tabular}}
% \vspace{-1em}
\end{table}

\subsubsection{Data Abundant}

Table \ref{ner_fully_result_table} shows the performance of each method evaluated on the full CoNLL2003 dataset.
As we can observe that our framework has the same performance on the F1 score as the best baseline LightNER, which is a prompt-based framework proposed for NER.
% Remarkably, compared to three finetuning-based methods, our CPGF-based NER model achieves higher F1 scores whether using T5-base or T5-large.
Remarkably, compared to three finetuning-based methods, CPGF achieves higher F1 scores, whether using T5-base or T5-large.
The experiment on complete CoNLL2003 shows that our framework is practical for the NER task in the data-abundant scenario.

\begin{table}[t]
\vspace{-1em}
\centering
\renewcommand\arraystretch{1}
\caption{Results (\%) for NER in low resource scenario. *indicates the low-resource entity type.}
\label{ner_low_resource_result_table}
\setlength{\tabcolsep}{3.3mm}{
\begin{tabular}{lccccc}
\toprule
Model   & PER & ORG & LOC* & MIS* & Overall \\ \hline
LC-BERT    & \multicolumn{1}{c}{76.25}    & \multicolumn{1}{c}{75.32} & 61.55 & 59.35 & 68.12 \\
LC-BART   & \multicolumn{1}{c}{75.70} & \multicolumn{1}{c}{73.59} & 58.70 & 57.30 & 66.82 \\ 
Template   & \multicolumn{1}{c}{84.49} & \multicolumn{1}{c}{72.61} & 71.98 & \textbf{73.37} & 75.59 \\ 
LightNER    & \multicolumn{1}{c}{90.96}    & \multicolumn{1}{c}{76.88}    & 81.57  & 52.08 & 78.97 \\ \hline
CPGF-base  & \multicolumn{1}{c}{95.92} & \multicolumn{1}{c}{80.28} & 80.91 & 70.62 & 82.90 \\
% base \textit{w/} hint    & \multicolumn{1}{c}{0} & \multicolumn{1}{c}{0} & 0  & 0 & 0\\
CPGF-large  & \multicolumn{1}{c}{\textbf{96.28}} & \multicolumn{1}{c}{\textbf{80.38}} & \textbf{84.12} & 70.40 & \textbf{84.80} \\
% large \textit{w/} hint    & \multicolumn{1}{c}{0} & \multicolumn{1}{c}{0} & 0 & 0 & 0\\
\bottomrule
\end{tabular}}
\vspace{-1em}
\end{table}

\subsubsection{Low Resource}

For comparison purposes, we followed the few-shot NER setting used in LightNER \cite{lightNER} that limits the number of instances for certain specific categories in the training set by downsampling on CoNLL2003.
Specifically, ``LOC'' and ``MIS'' are considered low-resource entities, while ``PER'' and ``ORG'' are set as high-resource entities.
In this scenario, we evaluated our CPGF-based NER models with the same training instances used in LightNER.

We can observe from Table \ref{ner_low_resource_result_table} that both CPGF-base and CPGF-large achieve remarkable increases in high-resource entity categories.
When it comes to low-resource entities, our CPGF-large makes a significant improvement for the type ``LOC'', and both the two CPGF-based models obtain competitive results on ``MIS''. 
It demonstrates that our framework can effectively identify entities in low-resource scenarios.

\subsection{Results for Relation Extraction}

\subsubsection{Baselines}
PTR \cite{PTR_2021} is the first work to introduce the prompt paradigm into the RE task.
Konwprompt \cite{knowprompt_2022} is the current state-of-the-art prompt-based RE approach that integrates the label information of relationships into the virtual tokens of prompts on its basis.
We selected both PTR and Konwprompt as baselines for the RE task.

\begin{table}[thb]
\vspace{-1em}
\centering
\renewcommand\arraystretch{1}
\caption{Results (\%) for RE on the complete SemEval dataset.}
\label{re_fully_result_table}
\setlength{\tabcolsep}{3.5mm}{
% \resizebox{1\columnwidth}{!}{

\begin{tabular}{lccc}
\toprule
Model   & P & R & F1 \\ \hline
PTR                   & \multicolumn{1}{c}{89.49} & \multicolumn{1}{c}{90.72} & 90.10 \\ 
KonwPrompt      & \multicolumn{1}{c}{-}    & \multicolumn{1}{c}{\textbf{-}}    & 90.20 \\ \hline
CPGF-base  & \multicolumn{1}{c}{\textbf{89.59}} & \multicolumn{1}{c}{90.58} & 90.09 \\
% base \textit{w/} hint    & \multicolumn{1}{c}{0} & \multicolumn{1}{c}{0} & 0  \\
CPGF-large  & \multicolumn{1}{c}{89.17} & \multicolumn{1}{c}{\textbf{91.33}} & \textbf{90.24} \\
% large \textit{w/} hint    & \multicolumn{1}{c}{0} & \multicolumn{1}{c}{0} & 0 \\
\bottomrule
\end{tabular}}
\vspace{-1em}
\end{table}

\begin{figure}[thb]
\vspace{-1em}
\centering
\includegraphics[width=1\columnwidth]{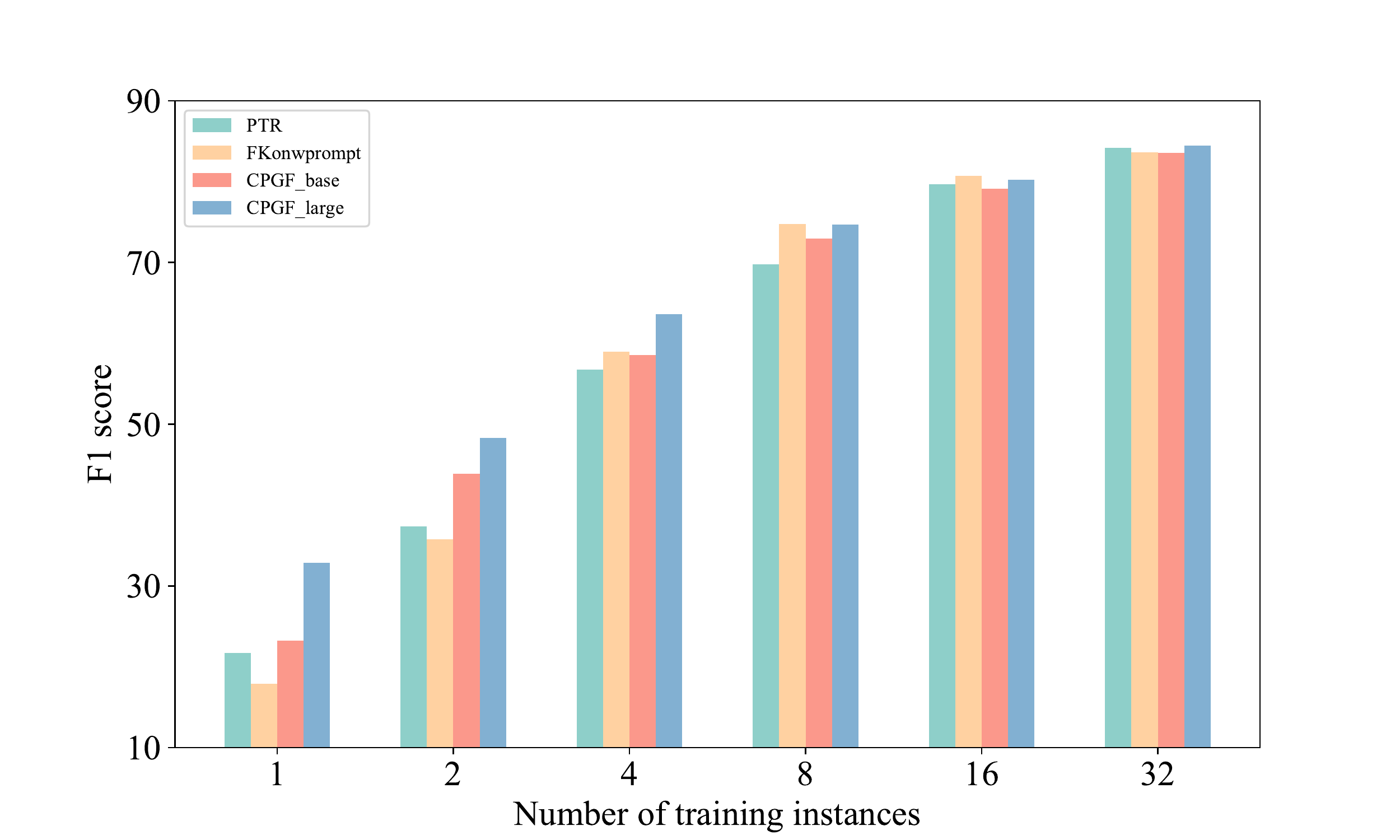} 
%\vspace{-1em}% Reduce the figure size so that it is slightly narrower than the column. Don't use precise values for figure width.This setup will avoid overfull boxes.
% \caption{An example of pre-training and fine-tuning processes. The upper part shows the pre-training process of BERT, and the lower part is the overview of an approach that detects events by applying the CRF algorithm and fine-tuning BERT.}
\vspace{-1em}
\caption{Performance in low-resource scenarios with different sizes of training instances.}
\label{fig:re_few_shot}
\vspace{-1em}
\end{figure}

% \begin{figure}[t]
% \centering
% \includegraphics[width=1\columnwidth]{figs/re_few_shot.pdf} 
% %\vspace{-1em}% Reduce the figure size so that it is slightly narrower than the column. Don't use precise values for figure width.This setup will avoid overfull boxes.
% % \caption{An example of pre-training and fine-tuning processes. The upper part shows the pre-training process of BERT, and the lower part is the overview of an approach that detects events by applying the CRF algorithm and fine-tuning BERT.}
% \caption{Result for RE in Low-Resource.}

% \label{fig:re_few_shot}
% \end{figure}

\subsubsection{Data Abundant}

As displayed in Table~\ref{re_fully_result_table}, CPGF achieves the highest F1 score when applying T5-large as the PLM. It demonstrates the effectiveness of our framework on the Relation Extraction task.

% \begin{table}[thb]
% \centering
% \renewcommand\arraystretch{1}
% \caption{Performance in low-resource scenarios with different sizes of training instances.}
% \label{re_few_shot_table}
% \setlength{\tabcolsep}{0.1mm}{
% % \resizebox{1\columnwidth}{!}{

% % \begin{tabular}{lcccccc}
% \begin{tabular}{lllllll}
% \toprule
% \toprule
% Model & K=1 & K=2 & K=4 & K=8 & K=16 & K=32  \\ \hline
% PTR      & 21.74 & 37.34 & 56.77 &  69.79  & 79.69 & 84.20 \\ 
% KonwPrompt     & 17.92 & 35.79 & 58.94 & \textbf{74.74} & \textbf{80.72} &  83.65 \\ \hline
% % CPGF-base & 23.20 & 43.87 & 58.52 & 72.96 & 79.13 & 83.56 \\
% % base \textit{w/} hint    & \multicolumn{1}{c}{0} & \multicolumn{1}{c}{0} & 0  \\
% CPGF-large & \textbf{32.84}\textcolor{red}{\tiny{(+14.92)}} & \textbf{48.30}\textcolor{red}{\tiny{(+12.51)}} & \textbf{63.61}\textcolor{red}{\tiny{(+4.67)}} & 74.67\textcolor{blue}{\tiny{(-0.07)}} & 80.23\textcolor{blue}{\tiny{(-0.49)}} & \textbf{84.44}\textcolor{red}{\tiny{(+0.79)}} \\
% % large \textit{w/} hint    & \multicolumn{1}{c}{0} & \multicolumn{1}{c}{0} & 0 \\
% \bottomrule
% \end{tabular}}
% \end{table}

\subsubsection{Low Resource}

In this scenario, we not only followed the previous works \cite{PTR_2021, knowprompt_2022} to conduct 8-, 16-, and 32-shot experiments but also extended three more extreme experimental groups: 1-, 2-, and 4-shot.
We maintained the same data sampling method and the metric of average performance as PTR and KnowPrompt.

The results of the above experimental groups are shown in Fig.\ref{fig:re_few_shot}.
CPGF achieves competitive results under the traditional 8-, 16-, and 32-shot settings while improving significantly in the rest of the experimental groups with fewer training instances.
Firstly, the data in the chart demonstrates that our framework can effectively extract relation information in low-resource scenarios.
Moreover, the remarkable improvement suggests that CPGF is more suitable for scenarios where training data is extremely scarce than existing prompt-based methods.

\begin{table*}[t]
\vspace{-1em}
\centering
\renewcommand\arraystretch{1.1}
\caption{Results (\%) for IE tasks with various prompts.}
\label{table:different_prompt}
\setlength{\tabcolsep}{2mm}{
% \resizebox{1\columnwidth}{!}{

\begin{tabular}{p{14cm}ccc}
\toprule
Prompt Text Example  & P & R & F1 \\ \midrule
\multicolumn{4}{c}{Named Entity Recognition} \\ \midrule
In the sentence above, words [mask0] indicates Locations.  & 92.24 & 92.37 & 92.31 \\ \hline 		
In the sentence above, words [mask0] indicates positions or sites occupied or available for occupancy or marked by some distinguishing feature.   & 92.57 & 91.52 & 92.04 \\ \hline
In the sentence above, words [mask0] indicates Locations, such as BRUSSELS, Germany and Britain. & 91.76 & 92.53 & 92.15 \\ \hline
Locations is(are) [mask0].     & 92.07    & 91.93    & 92.00 \\ \midrule

\multicolumn{4}{c}{Event Extraction (Trigger Detection)} \\ \midrule
There is an event with the type demonstrate triggered by word [mask0]. In this event ...  & 75.34 & 79.62 & 77.42 \\ \hline
In the above sentence, the key word [mask0] indicates that an event with the type demonstrate occurs. ... & 76.31    & 78.01    & 77.15 \\ \hline
% base \textit{w/} hint    & \multicolumn{1}{c}{0} & \multicolumn{1}{c}{0} & 0  \\
In the above sentence, word [mask0] is the trigger word for a demonstrate event. ... & 76.09 & 77.85 & 76.95 \\ \hline
In the above sentence, word [mask0] indicates a demonstrate event. ...  & 75.96 & 77.89 & 76.91 \\ \midrule

\multicolumn{4}{c}{Relation Extraction} \\ \midrule
From the above sentence, the following conclusion can be inferred: it is [mask0] that the organization ``Retired military officers'' is located in the country ``Malaysia''.       & 89.59 & 90.58 & 90.09 \\ \hline
From the above sentence we can know that, it is [mask0] that the organization ``Retired military officers'' is located in the country ``Malaysia''.     & 89.48    & 89.83    & 89.65 \\ \hline
So the inference ``the organization Retired military officers is located in the country Malaysia'' is [mask0]. & 88.78 & 90.72 & 89.72 \\ \hline
% base \textit{w/} hint    & \multicolumn{1}{c}{0} & \multicolumn{1}{c}{0} & 0  \\
That is to say , it is [mask0] that the organization ``Retired military officers'' is located in the country ``Malaysia''.  & 89.37 & 90.32 & 89.84 \\ 
\bottomrule
\end{tabular}}
\vspace{-1em}
\end{table*}

\begin{table*}[thb]
\centering
\renewcommand\arraystretch{1.1}
\caption{Results (\%) for RE task with different template designs.}
\label{table:re_different_template}
\setlength{\tabcolsep}{2mm}{
% \resizebox{1\columnwidth}{!}{

\begin{tabular}{lp{10cm}cc}
\toprule
Template name & Example & Answer & F1 \\ \hline
Template recommended & From the above sentence, the following conclusion can be inferred: it is [mask0] that the organization ``Retired military officers'' is located in the country ``Malaysia''. & right / wrong     & 90.09 \\ \hline
Predicting consistency  & The above sentence is [mask0] with `` the organization Retired military officers is located in the country Malaysia ''.  & consistent / inconsistent   & 89.38
\\ \hline
Predicting head entity  & That is to say, the organization [mask0] is located in the country Malaysia. & span of the head entity & 83.54 \\ \hline
Predicting tail entity  & That is to say, the organization Retired military officers is located in the country [mask0]. & span of the tail entity & 79.44 \\ \hline
Predicting relation  & From the above sentence, the following conclusion can be inferred: the relationship between Retired military officers and Malaysia is [mask0]. & name of the relation  & 81.26 \\
\bottomrule
\end{tabular}}
\end{table*}

\subsection{Discussion}

\subsubsection{Effect of prompts' text composition}

Existing studies \cite{do_prompt_,prompt_shunxu} have shown that the form of prompts has a non-negligible impact on the performance of prompt-based methods.
According to the report in the paper \cite{template_BART}, the performance fluctuation of the NER system can even reach 20\% just by changing the textual representation of the prompt.
We designed multiple different manual prompts for each IE task to investigate the impact of the prompt text composition in CPGF.
The PLM used in this experiment is T5-base.
As we can see from Table~\ref{table:different_prompt}, the change of prompt text causes perturbations within a narrow range.
However, these small fluctuations are negligible compared to the
overall scores.
This suggests that CPGF is slightly affected by the prompt text. 
In other words, our framework can reduce the labor cost of prompt design engineering than other prompt-based methods.

\subsubsection{Impact of Converted Form for RE}

Besides the form of judging semantic consistencies, there are other ways to extract relations from text using CPGF.
To discuss the effectiveness of our RE approach, we compared it with methods as follows:
\begin{itemize}
    \item \textbf{Predicting consistency} is similar to the form we recommend but directly predicts ``consistent'' or ``inconsistent'' rather than ``right'' or ``wrong''. 
    \item \textbf{Predicting head entity} determines whether the relationship exists by predicting the head entity in the entity pair.
    Specifically, given the tail entity and description of the target relationship, the relation is considered to exist only if the model accurately predicts the head entity span.
    \item \textbf{Predicting tail entity} shares the same idea with \textbf{Predicting head entity}, but the tail entity is what needs to be predicted.
    \item \textbf{Predicting relation} aims to output the name of the relation type directly. 
\end{itemize}
We implemented the above methods with the same experimental setup (using T5-base) and presented their performances in Table~\ref{table:re_different_template}.
Experimental results show that transforming the RE task into the form of judging semantic consistencies outperforms other forms significantly.
Compared to directly outputting ``consistent'' or ``inconsistent'', the form we recommend achieves a higher score.

% \begin{figure*}[!t]
% \centering
% \vspace{-1em}
% \subfloat[]{\includegraphics[width=2.36in]{figs/ner-lr-22.pdf}%
% % \label{fig_first_case}
% }
% \hfil
% \subfloat[]{\includegraphics[width=2.36in]{figs/ner-wd-22.pdf}%
% % \label{fig_second_case}
% }
% \hfil
% \subfloat[]{\includegraphics[width=2.36in]{figs/ner-bs-22.pdf}%
% % \label{fig_third_case}
% }
% \hfil
% \vspace{-1.25em}
% \subfloat[]{\includegraphics[width=2.36in]{figs/re-lr-22.pdf}%
% % \label{fig_first_case}
% }
% \hfil
% \subfloat[]{\includegraphics[width=2.36in]{figs/re-wd-22.pdf}%
% % \label{fig_second_case}
% }
% \hfil
% \subfloat[]{\includegraphics[width=2.36in]{figs/re-bs-22.pdf}%
% % \label{fig_third_case}
% }
% \caption{Changes in F1 scores (\%) with tuning hyper-parameters. (a), (b) and (c) are results on span extraction. (d), (e) and (f) are results on relation extraction.}
% \vspace{-1em}
% \label{fig:para}
% \end{figure*}

\begin{figure*}[!t]
\centering
\vspace{-1em}

\vspace{-1.25em}
\subfloat[]{\includegraphics[width=2.36in]{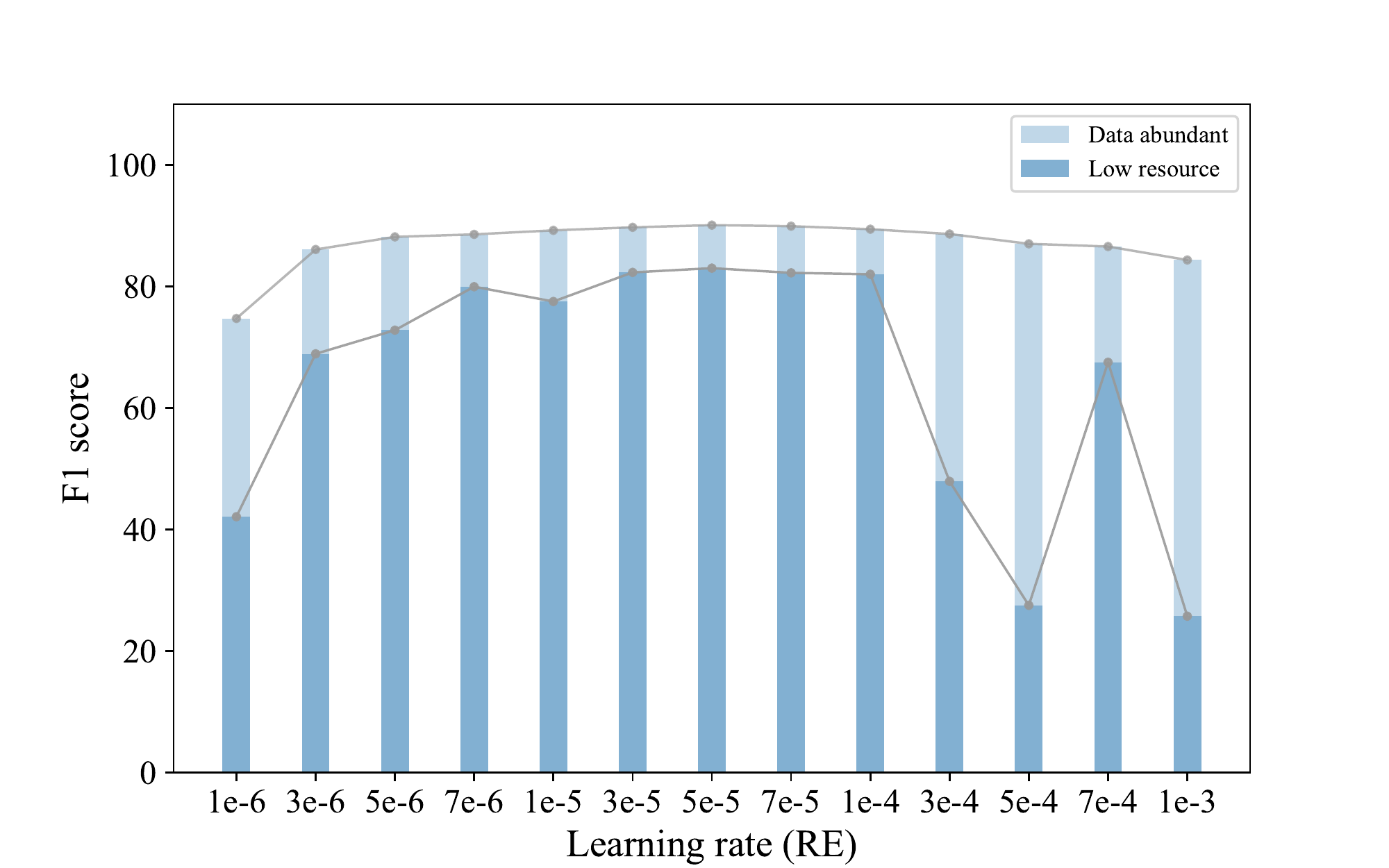}%
% \label{fig_first_case}
}
\hfil
\subfloat[]{\includegraphics[width=2.36in]{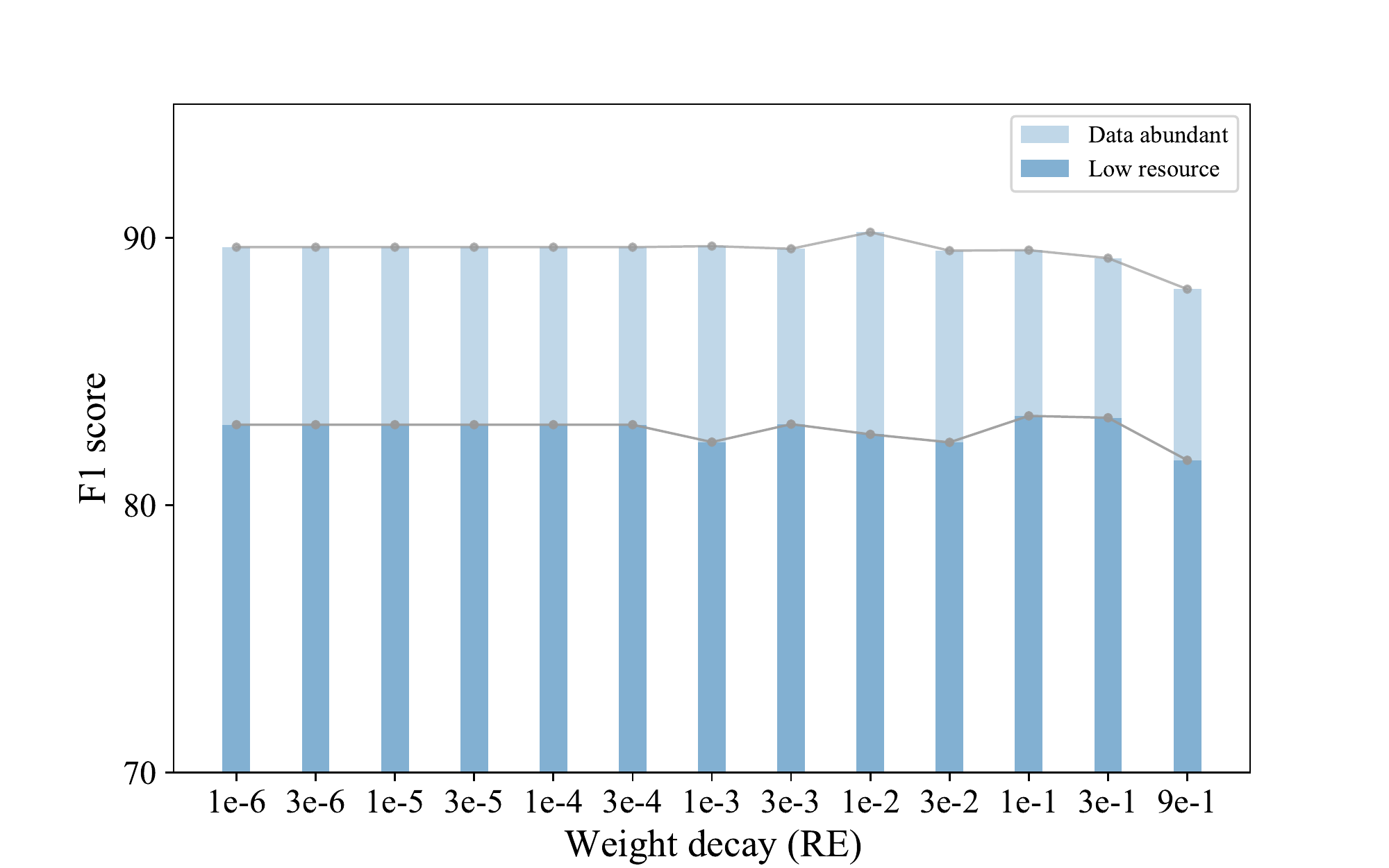}%
% \label{fig_second_case}
}
\hfil
\subfloat[]{\includegraphics[width=2.36in]{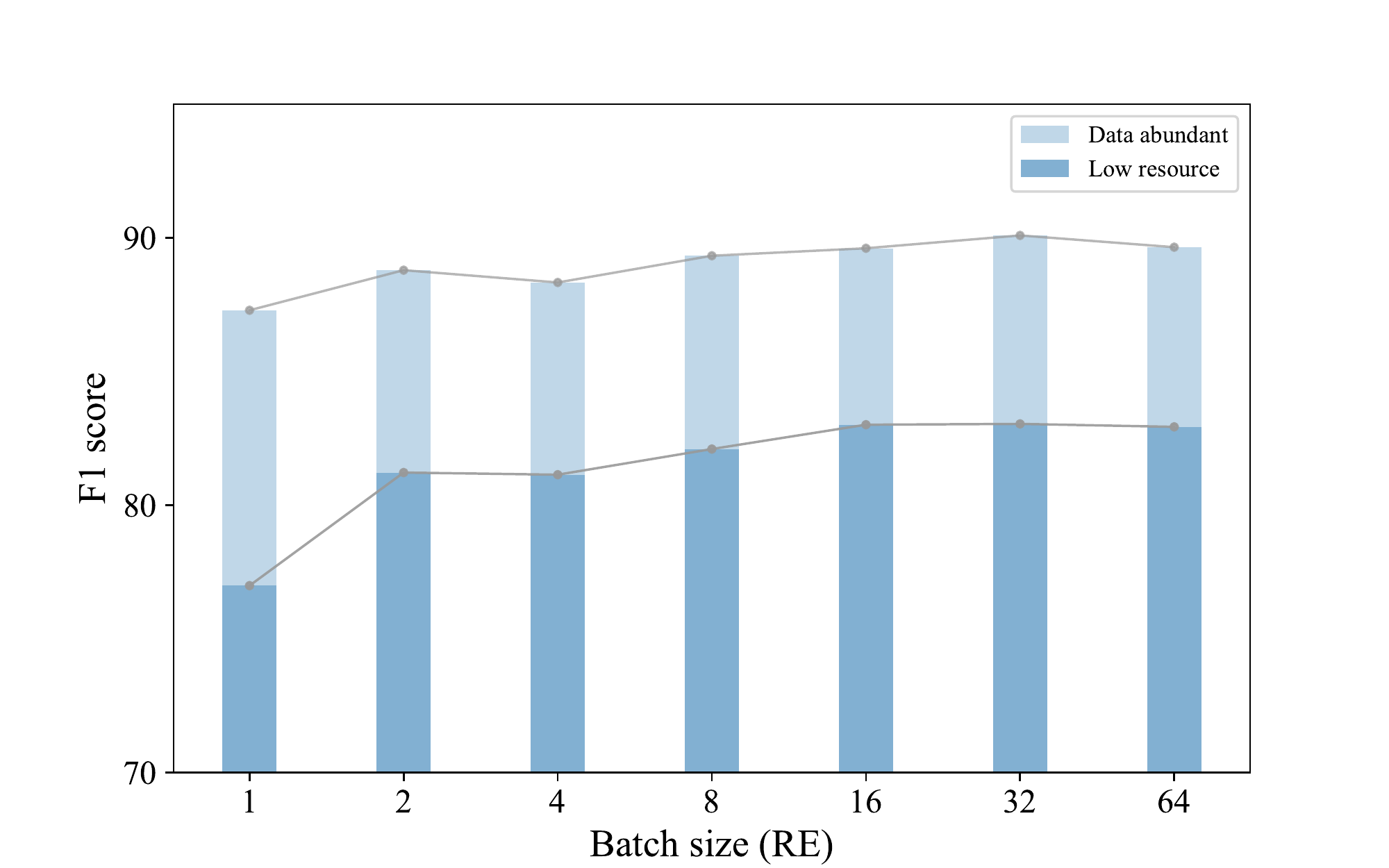}%
% \label{fig_third_case}
}
\hfil
\subfloat[]{\includegraphics[width=2.36in]{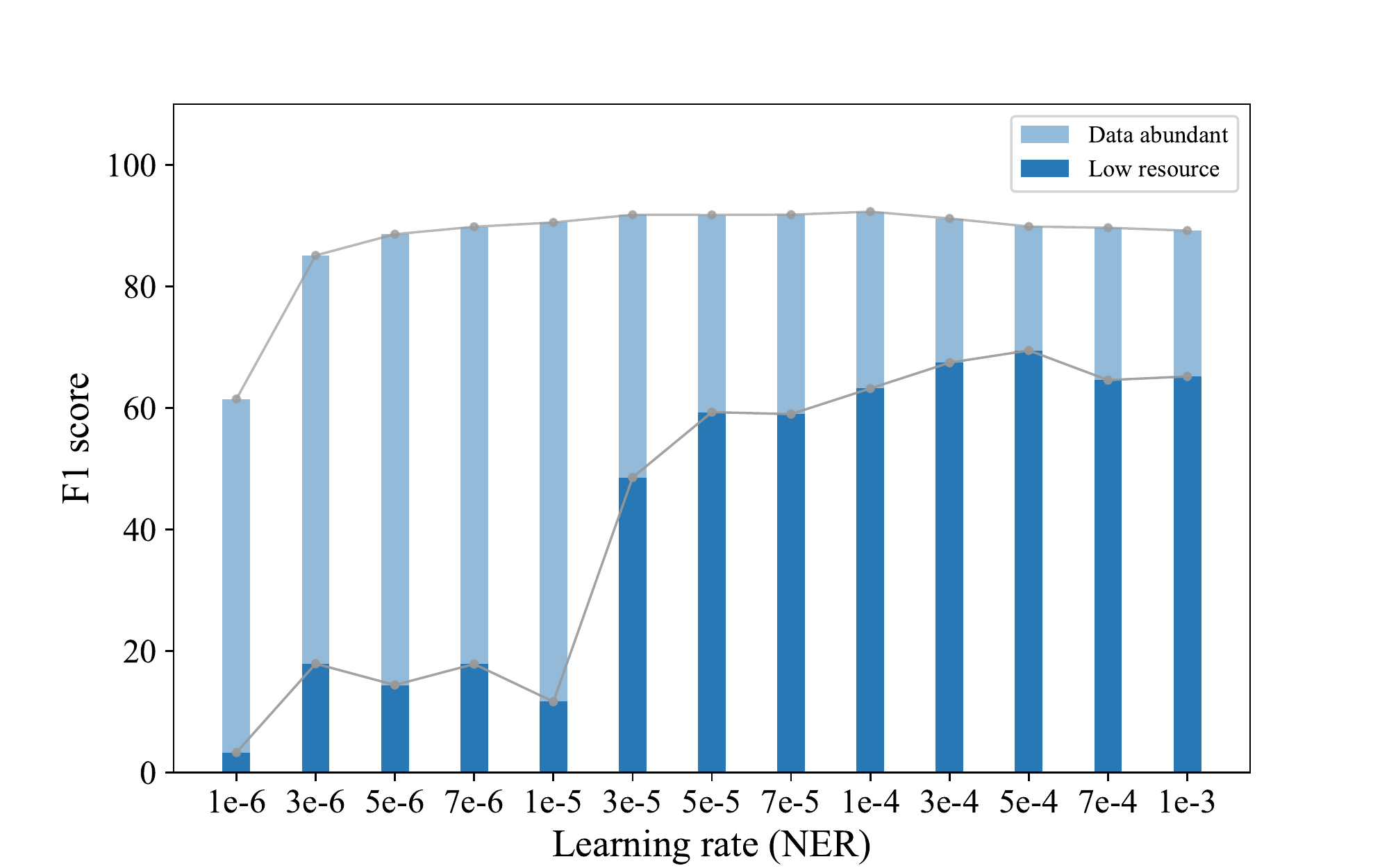}%
% \label{fig_first_case}
}
\hfil
\subfloat[]{\includegraphics[width=2.36in]{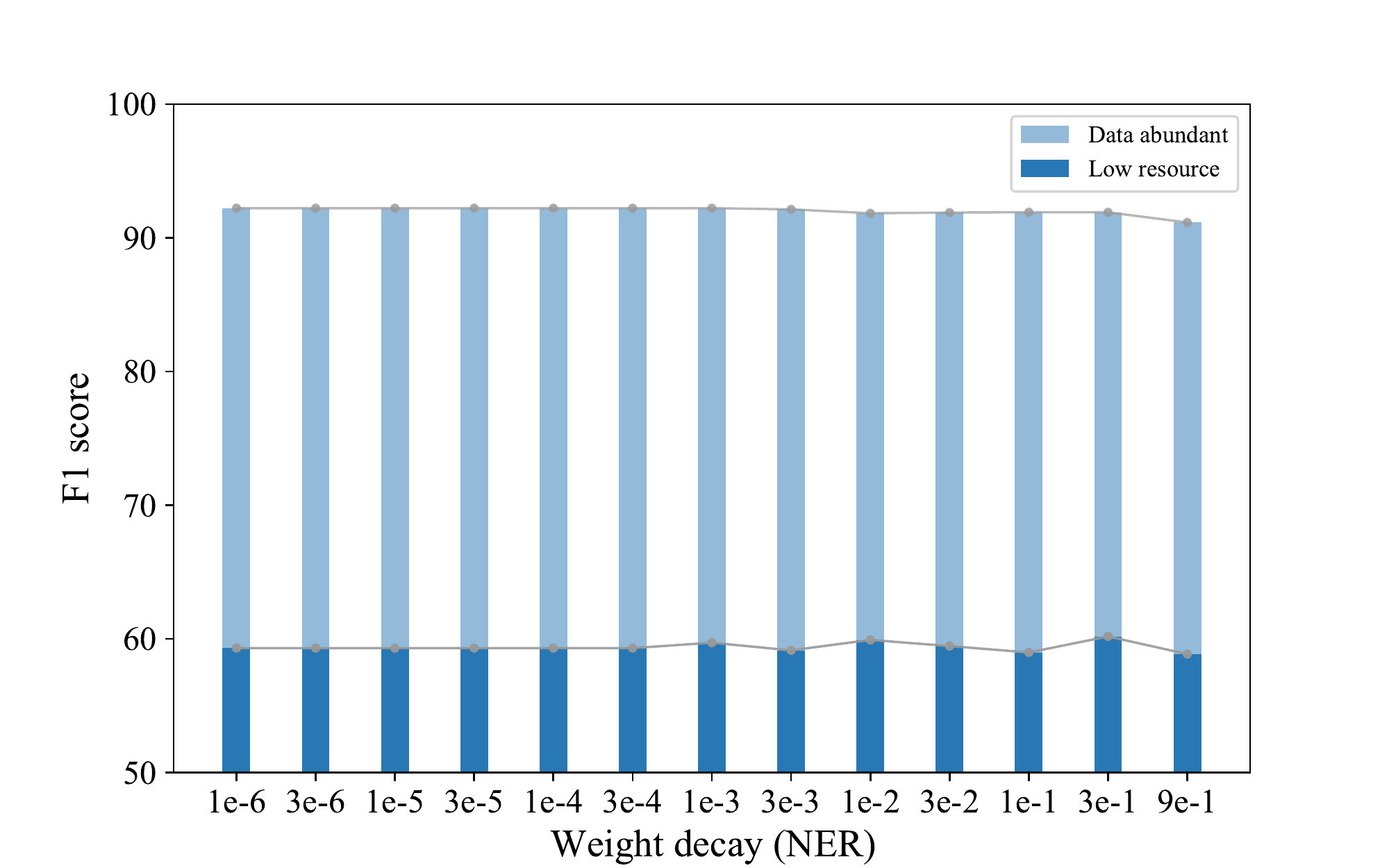}%
% \label{fig_second_case}
}
\hfil
\subfloat[]{\includegraphics[width=2.36in]{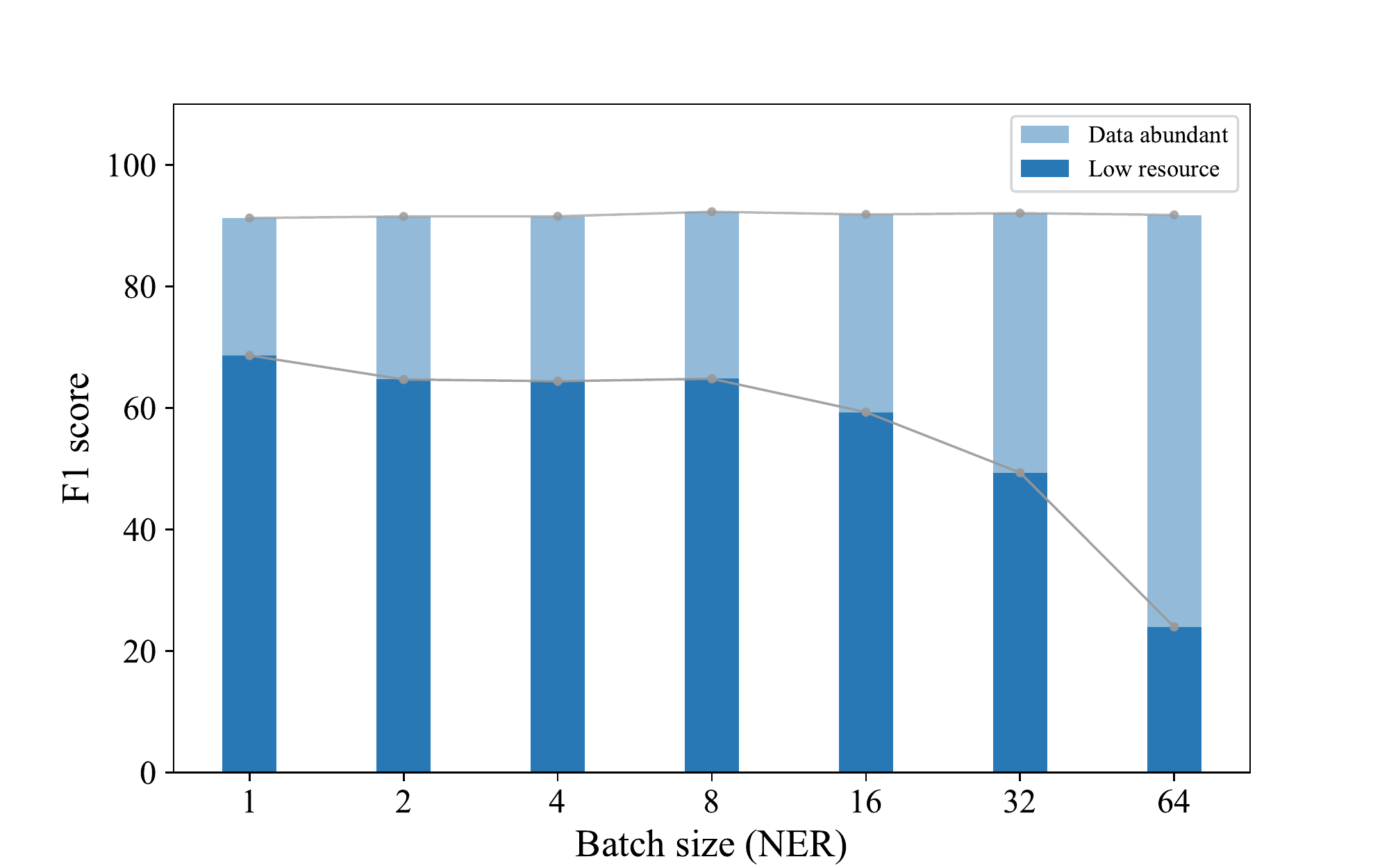}%
% \label{fig_third_case}
}

\caption{Changes in F1 scores (\%) with tuning hyper-parameters. (a), (b) and (c) are results on Relation Extraction, while (d), (e) and (f) are results on span extraction. The performance in the low-resource scenario is more affected by hyper-parameters than in the data-abundant scenario. The learning rate and batch size values significantly affect the framework's performance on various tasks.}
% \vspace{-1em}
\label{fig:para}
\end{figure*}

\subsubsection{Effect of Hyper-parameters}

Selecting hyper-parameters is a very important process for training neural networks.
In this subsection, we study the effect of each hyperparameter on CPGF in different scenarios.
As can be seen from Table~\ref{parameters_table}, NER shares the same hyperparameter settings with EE, while RE does not.
Because the purposes of NER and EE are to extract spans, which differ from RE.
Therefore we refer to NER and EE collectively as span extraction.
We take NER as an example to show the effect of hyperparameter settings on span extraction tasks.
Experiments on data-abundant scenarios were conducted on completed datasets. 
In low-resource scenarios, we replaced training and validation sets with 32 samples which are randomly sampled from them.
Experiments were conducted on T5-base.

As we can observe from Fig.\ref{fig:para}, learning rate changes significantly impact both tasks' results in the low-resource scenario.
In the scenario with sufficient data, the magnitude of this effect becomes slight.
But setting learning rates to too small values will also seriously affect performance.
Compared to learning rates, values of weight decay have milder effects on tasks' results in all scenarios. 
However, selecting an appropriate weight decay value will allow CPGF to perform better.

With the increase in batch size, the performance trends of these two tasks are different.
For the RE task, the effect of batch size in the two scenarios has same regularities: 1) changing batch size has a noticeable but modest effect on the F1 score, and 2) the optimal value is 32.
When applying CPGF to span extraction tasks, these consistent regularities are no longer present.
In the data-abundant scenario, batch size has a very slight impact on performance.
Nevertheless, in the low-resource scenario, increasing batch size significantly drops the F1 score.
We believe that differences in task form lead to this phenomenon.
When extracting relations, the output is one of the two fixed words, "right" and "wrong."
On the contrary, CPGF predicts "[mask]" tokens as variable-length phrases with various expressions to extract entities or events.
For the latter task, the model will find the gradient descent direction faster and more accurately with a smaller batch size in the low-resource scenario.

% As we can observe from Fig.\ref{fig:para}, in the data abundant scenario, the impact of weight decay and batch size on both NER and EE is slight.
% But setting the learning rate to a too small value can seriously affect performance.
% In the scenario of low source, changes in both A and B can cause dramatic fluctuations in performance. 
% The figure suggests that CPGF tends to learn knowledge from a few samples with a large learning rate and small batch size.
% Weight decay has a certain effect on the results, but it is less obvious than the other two.

\subsubsection{Apply Prompt Templates to Other PLMs}

% \begin{table}[h]
% \centering
% \renewcommand\arraystretch{1}
% \caption{Results (\%) for RE on different PLMs.}
% \label{table:re_PLMs}
% \setlength{\tabcolsep}{3.5mm}{
% % \resizebox{1\columnwidth}{!}{

% \begin{tabular}{lccc}
% \toprule
% PLM   & P & R & F1 \\ \hline
% RoBERTa-base                   & \multicolumn{1}{c}{88.61} & \multicolumn{1}{c}{89.39} & 89.00 \\ 
% RoBERTa-large      & \multicolumn{1}{c}{86.20}    & \multicolumn{1}{c}{\textbf{92.22}}    & 89.11 \\ \hline
% T5-base  & \multicolumn{1}{c}{\textbf{89.59}} & \multicolumn{1}{c}{90.58} & 90.09 \\
% % base \textit{w/} hint    & \multicolumn{1}{c}{0} & \multicolumn{1}{c}{0} & 0  \\
% T5-large  & \multicolumn{1}{c}{89.17} & \multicolumn{1}{c}{\textbf91.33} & \textbf{90.24} \\
% % large \textit{w/} hint    & \multicolumn{1}{c}{0} & \multicolumn{1}{c}{0} & 0 \\
% \bottomrule
% \end{tabular}}
% \end{table}

\begin{figure}[thb]
\vspace{-1em}
\centering
\includegraphics[width=1\columnwidth]{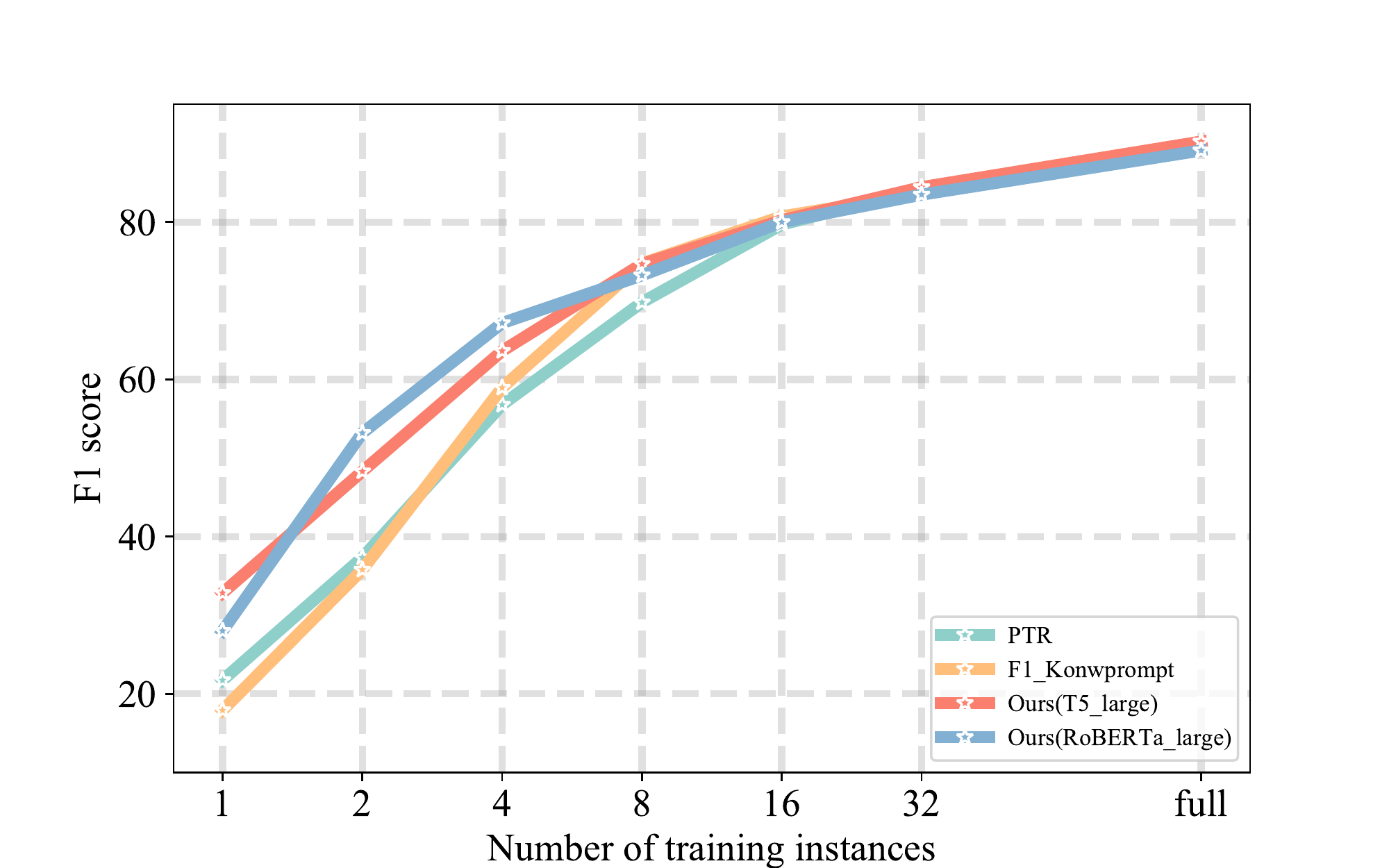} 
%\vspace{-1em}% Reduce the figure size so that it is slightly narrower than the column. Don't use precise values for figure width.This setup will avoid overfull boxes.
% \caption{An example of pre-training and fine-tuning processes. The upper part shows the pre-training process of BERT, and the lower part is the overview of an approach that detects events by applying the CRF algorithm and fine-tuning BERT.}
\vspace{-1em}
\caption{Performances for RE with different PLM.}
\label{fig:re_roberta}
\vspace{-1em}
\end{figure}

The task forms for NER and EE in this paper depend on PLMs with encoder-decoder architecture, such as T5 and BART.
The reason is that these two tasks require PLMs with the generative ability to output variable-length information.
Since our templates for NER and EE are designed based on the pre-training task of T5, they cannot be directly applied to BART.

The RE prompt template we designed in this paper can directly apply to encoder-only PLMs. 
We selected RoBERTa as the PLM to verify the effectiveness of our RE template on encoder-only PLMs.
Fig.~\ref{fig:re_roberta} shows the results in both data-abundant and low-resource scenarios.
It suggests that extracting relations by judging semantic consistencies is effective on RoBERTa.
Especially in situations where training instances are incredibly scarce, our method significantly outperforms baselines remarkably.
The improvements demonstrate the template we proposed for RE has better generalization ability on other PLMs.

\section{Conclusion}

This paper proposes a unified generative framework for Information Extraction tasks based on the prompt paradigm. 
It extracts information by predicting masked words in pre-designed prompts, which consist of one or multiple sub-prompts.
Meanwhile, a kind of composable prompt composed of modular sub-prompts is introduced to improve the model's generalization ability for complex information in the scenario of data-scarce.
Furthermore, we propose a novel approach that extracts relation types by estimating semantic inconsistencies in prompts.
Experiments conducted on Named Entity Recognition, Event Extraction, and Relation Extraction demonstrate the effectiveness of our framework in scenarios of data-abundant and data-scarce.
Impact factors of performances on various tasks are discussed through numerical experiments.

\bibliographystyle{IEEEtran}
\bibliography{IEEEref}

% \newpage


\newpage
~~
\vfill

\end{document}